\documentclass[journal]{IEEEtran}
\newcommand{\userdefinelength}{\linewidth}

\newcounter{mytempeqncnt}

\ifCLASSINFOpdf
\else
\fi
\usepackage{graphics,amssymb,graphicx, theorem,amsmath,amsfonts,epsfig,float,times,multirow, setspace, latexsym,psfrag,cite,ulem, algpseudocode, algorithm, amsmath }

\usepackage{xcolor}
\usepackage{soul}

\usepackage{subcaption}
\usepackage{graphicx}
\usepackage{lipsum}
\usepackage{siunitx}
\usepackage{placeins}

\allowdisplaybreaks

\newcommand{\be}{\begin{equation}}
\newcommand{\ee}{\end{equation}}
\newcommand{\ben}{\begin{eqnarray}}
\newcommand{\een}{\end{eqnarray}}
\newtheorem{theorem}{Theorem}
\newtheorem{lemma}{Lemma}

\newtheorem{corollary}{Corollary}

\begin{document}

\title{User-Number Threshold-based Base Station On/Off Control for Maximizing Coverage Probability}

\author{Jung-Hoon Noh~\IEEEmembership{Member,~IEEE}, and Seong-Jun Oh~\IEEEmembership{Senior,~IEEE}}
\date{\today}
\maketitle \thispagestyle{empty}

\begin{abstract}

In this study, we investigate the operation of user-number threshold-based base station (BS)   on/off control, in which the BS turns off when the number of active users is less than a specific   threshold value. This paper presents a space-based analysis of the BS on/off control system to which a stochastic geometric approach is applied. In particular, we derive the approximated closed-form expression of the coverage probability of a homogeneous network (HomNet) with the user-number threshold-based on/off control. Moreover, the optimal user-number threshold for maximizing the coverage probability is analytically derived. In addition to HomNet, we also derive the overall coverage probability and the optimal user-number thresholds for a heterogeneous network (HetNet). The results show that HetNet, the analysis of which seems intractable, can be analyzed in the form of a linear combination of HomNets with weighted densities. In addition, the optimal user-number threshold of each tier is obtained independently of other tiers. The modeling and analysis presented in this paper are not only limited to the case of user-number threshold-based on/off control, but also applicable to other novel on/off controls with minor modifications. Finally, by comparing with the simulated results, the theoretical contributions of this study are validated.

\end{abstract}



\section{Introduction}
\label{sec:intro}

One of the major trends in recent networks is its densification with the deployment of large numbers of base stations~(BSs)~\cite{Feng:2017}. 
In mmWave networks, for example, small-cell networks, distributed antenna systems, 
and large-scale MIMO networks, BSs are deployed close to users to cope with propagation loss to improve the signal-to-noise ratio (SNR) and to reduce service delay. 
However, these benefits impose high costs, such as a significant increase in the total energy consumption of wireless systems due to large-scale BS deployments. 
High energy consumption not only increases the cost of wireless operators, but also increases greenhouse gas emissions~\cite{Hasan:2011}.

\subsection{Related Works}

As a useful technique to reduce energy consumption, BS on/off control schemes have been introduced in several studies to date~\cite{Feng:2017,McLaughlin:2011,Son:2011,Soh:2013,Han:2016,Liu:2016,Mugume:2019,Guo:2016,3GPP_TR:2013,Yu:2014,Park:2016,Park:2020}. 
The primary goal of such schemes is to reduce energy consumption by turning off certain BSs. 
Studies have shown that traffic demand varies greatly over space and time~\cite{Marsan:2009,Auer:2011}.
However, current cellular networks usually assume that the traffic demand is always high for all BSs,
and they are designed to support maximum traffic all the time.
From this point of view, there is a potential for energy saving by adopting on/off control of underutilized BSs. In addition to these, turning off certain BSs can have an additional effect of reducing overall inter-cell interference. 

Meanwhile, 
even if the overall inter-cell interference is reduced, 
users located in the coverage of an inactive BS connect to a BS located farther away, 
resulting in poor link quality. In the case of interference-limited scenarios, 
it is widely believed that these two effects are equally offset by each other, 
and the coverage probability is independent of the density of the BSs~\cite{Jeffrey:2011}. 
This result also applies to a heterogeneous network (HetNet)~\cite{Dhillon:2012}; however, it is only valid for the random on/off control scheme.

By employing coordinated on/off control rather than a random control scheme, we can make one effect prevail over the other effect; that is, inter-cell interference mitigation effects can prevail over the adverse effect of the increased distance from the user to the BS. 
Thus, coordinated on/off control can increase the signal-to-interference-plus-noise ratio (SINR) of the typical user while reducing the overall power consumption. 
Therefore, it is natural that employing such a scheme would increase energy efficiency with respect to that of random on/off methods. 
In \cite{Soh:2013}, a sleep strategy is introduced that relies on the activities of the BS by defining an arbitrary function for the activity levels of each BS. 
However, the arbitrary activity level function does not reflect the practical activity model, such as the traffic load or the number of served users. 

As a practical candidate for coordinated on/off control, strategies based on the number of users have been proposed in various studies. 
In \cite{3GPP_TR:2013}, a dynamic small-cell on/off scheme was proposed, where BSs are assumed to turn on or off instantaneously based on the user arrival or departure. 
To reduce the number of mode transitions, \cite{Yu:2014} proposed a dual-threshold scheme in which BSs turn on when the number of active users reaches a certain threshold and turn off when the number of active users reaches zero. 
\cite{Park:2016} proposed a distributed iterative threshold control method that computes the user-number threshold in a distributed manner to reduce the complexity. 
In \cite{Park:2020}, analytical formulas for various performance metrics and the optimal user-number threshold are presented for small-cell on/off control schemes. 
In these works, analyses are performed from the perspective of the time domain to model users' arrival with a Poisson process in which the active-mode service time is exponentially distributed. 
Few studies of user-number-based BS on/off control in terms of the space domain have been conducted so far. 
\cite{Liu:2016} presented a strategic sleeping approach based on the number of users in the coverage area and the distance between the BS and the user. 
However, for the strategic sleeping, no analysis results were presented, but only for a random sleeping. 

\subsection{Paper Overview}
In this study, a space-based analysis of networks with user-number threshold-based on/off control is performed, which is the first such attempt to the best of the authors' knowledge. 
We take a snapshot of the network and apply tools from stochastic geometry to analyze the performance metrics of cellular networks with the deployment of user-number threshold-based on/off control. 
The stochastic geometry has been used extensively as an analytical tool to study cellular networks with improved tractability~\cite{Jeffrey:2011, Haenggi:2012, ElSawy:2017}. 
In particular, we model the distribution of BSs and users as an independent Poisson point process~(PPP).


One question we will explore throughout this study is the effect of the user-number threshold on the coverage probability. 
As the user-number threshold increases, the number of BSs in the off-mode increases, so the total amount of interference decreases.
However, the users in the off-mode BS connect to the BS located further away. Thus they are likely to have a lower SINR despite the reduced interference. Then, as the number of BSs in the off-mode increase, such users increase and their distance to the serving BS increases as well, worsening their link quality.
Depending on the user-number threshold, the impact of these two conflicting effects vary, and so does the overall coverage capability.


To answer this question, we derive the overall coverage probability of the network when employing user-number threshold-based on/off control, which represents the major theoretical contribution of this work. 
The analysis of a homogeneous network (HomNet) is performed first. 
Owing to the significant complexity, the coverage probability for the HomNet is approximated. 
Then, we formulate the optimization problem of finding the optimal user-number threshold for maximizing the overall coverage probability. 
The optimal user-number threshold proves to be the integer closest to the ratio of BS and user density.

In addition to the HomNet analysis, we also investigate user-number threshold-based on/off control in a HetNet consisting of multiple tiers of different transmit power and density. 
The analysis of the HetNet is significantly challenging because the coverage area of each tier is dependent on the parameters of all tiers in the network, i.e., density and transmit power. 
However, we prove that each tier in the HetNet is equivalent to a HomNet with weighted density, 
in which the weight is the inverse of the probability that a typical user connects to the BS of each tier. 
Thus, the analytical results of the HomNet can be extended to the HetNet. 
Furthermore, we obtain the optimal user-number thresholds, which maximize the overall coverage probability. 
It is shown that the solution is quite simple, because the optimal threshold for each tier is independent of the other tiers. 
Finally, all the analytical results are verified by simulation results.

\subsection{Major Contribution}
The contributions of this study are summarized as follows. 
First, this work provides general analytical tools and frameworks that can be used to evaluate the performance of various coordinated on/off controls. 
The coverage probability expression shown in this work is not limited to only user-number threshold-based on/off control, 
but can be extended to other state-of-the-art on/off control approaches.

Moreover, we obtain insight into the optimal on/off criteria for maximizing the coverage probability in the various coordinated on/off controls. 
From the optimal user-number threshold results, it can be assumed that the optimum on/off criterion for general coordinated on/off control is near the average level of BS activity.

Last but not least, we show that the analysis of the HetNet can be greatly simplified through an approach that equates the HetNet to a linear combination of HomNets by viewing each tier of the HetNet as a HomNet with the weighted density of the tier. 
This approach can also be applied to other performance metrics and problems based on stochastic geometry analysis, 
beyond the coverage probability for BS on/off control scheme. 

The remainder of this paper is organized as follows. In Section \ref{sec:sys_model}, we present our system model. In Section III, we analyze a HomNet with user-number threshold-based on/off control. In Section IV, we extend these results to a $K$-tier HetNet. In Section V, we validate our analysis through simulation results, and the paper is concluded in Section~VI.

\ifdefined\singlecolumn
\else
\begin{figure*}[htbp]
  \centering
  \begin{subfigure}[b]{0.49\linewidth}
  \vskip 0pt
  \includegraphics[width=\linewidth]{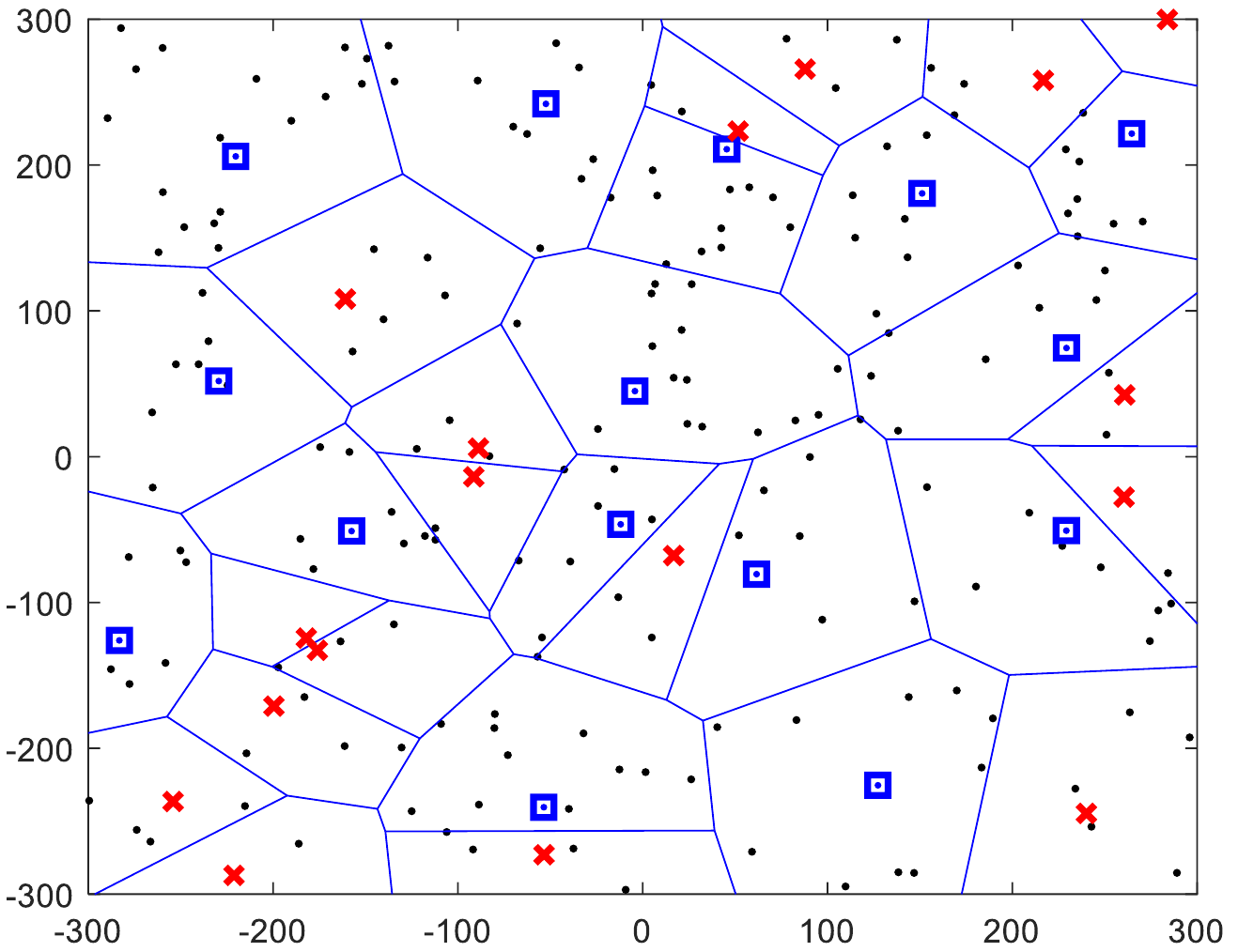}
  \vspace{3pt}
  \caption{Original cell boundaries\\ $\quad$ }
  \label{fig:BS_MS_distribution}
  \end{subfigure}
  \hfill
  \begin{subfigure}[b]{0.49\linewidth}
   \includegraphics[width=\linewidth]{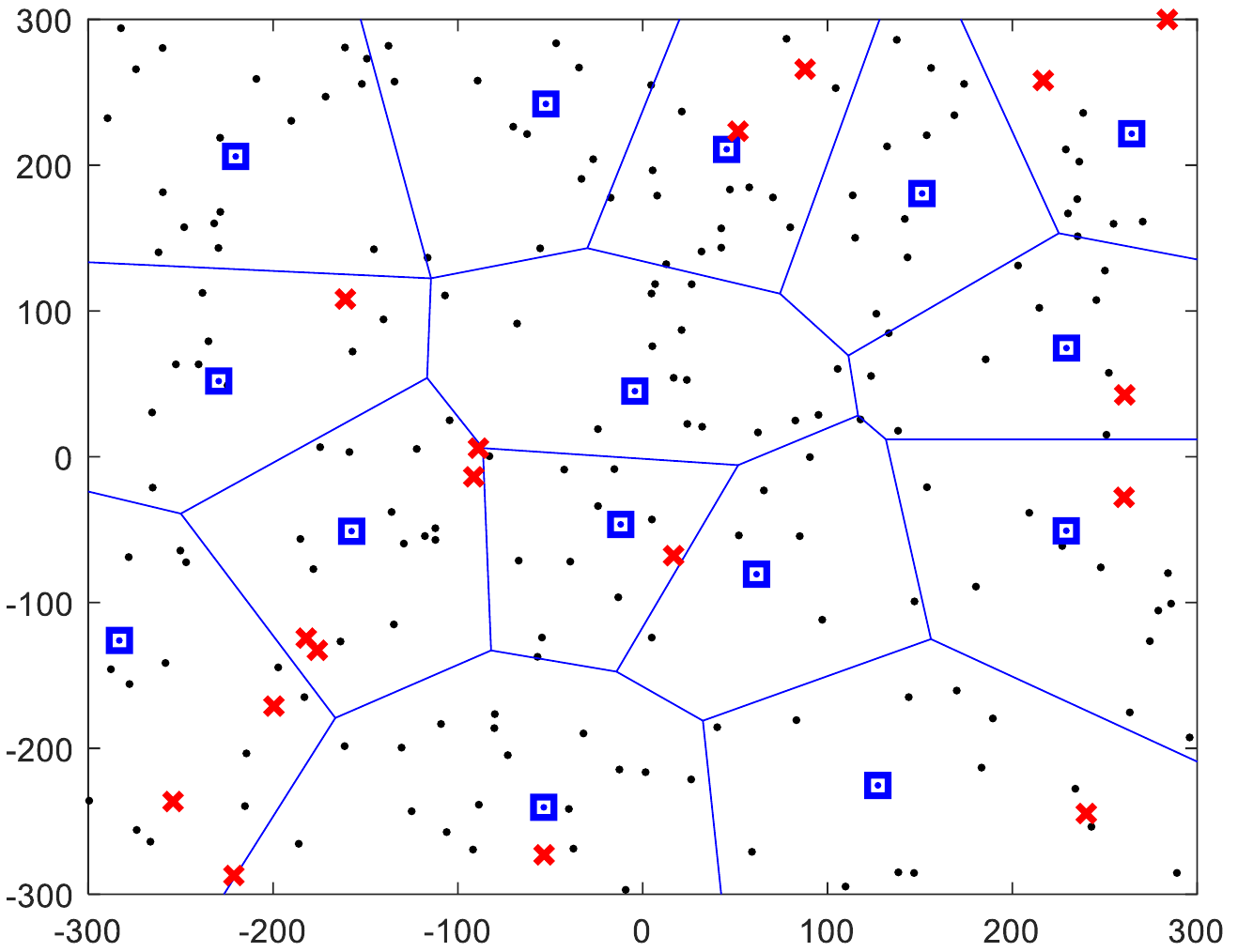}
  \vspace{3pt}
  \caption{Redrawn cell boundaries with user-number threshold-based on/off control where the user-number threshold $\theta$ is 5}
  \label{fig:BS_MS_distribution_w_onoff} 
  \end{subfigure}
  \caption{Poisson distributed BSs and users. Each user is associated with the nearest BS, where densities of BS and user are $\lambda_b=\num{1e-4}$ and $\lambda_u=\num{5e-4}$, respectively. The black dots are mobile users; the blue squares are the BS where the active number of users is above $\theta$; the red `x's are the BS where the active number of users is less than equal to $\theta$. The cell boundaries are shown and form a Voronoi tessellation.}
  \label{fig:voronoi_example}
  \vspace{-9pt}
\end{figure*}
\fi

\section{System Model}
\label{sec:sys_model}

\subsection{Network Model}

Consider a PPP-based HomNet in which BSs are located according to independent homogeneous PPP~$\Phi_b$ of density $\lambda_b$ in the Euclidean plane, 
and assume that all BSs transmit with power $P_t$.
The HetNet is modeled as a  $K$-tier cellular network in which each tier models a particular class of BS, such as a femtocell or picocell. Depending on the tier, BSs may differ in transmit power and spatial density. 
We assume that the BSs in the $i$-tier are spatially distributed as a PPP $\Phi_{i}$ of density $\lambda_{i}$, and transmit at power $P_{t,i}$.
The target SINR of all BSs in the network is assumed to be the same as $T$ for both the HomNet and HetNet. 
The mobile users are also modeled by an independent PPP $\Phi_u$ of density $\lambda_u$. 
We assume universal frequency reuse among the BSs.

In addition, radio resources are assumed to be allocated to multiple users within the coverage area by some form of orthogonal resource sharing (e.g., frequency or time division multiple access). 
The limited number of radio resources can be considered with multiple users under the blocking ratio constraint. 
However, such considerations entail significantly increased complexity, and are left to future work. 
In this study, the number of resources is assumed to be sufficiently large to accommodate all connected users.

\subsection{BS On/Off Control Scheme}
\label{sec:on_off_strategies}

The most common and straightforward strategy is random on/off control. 
Thus, its performance can be adopted as the baseline performance, 
which can be compared with that of more sophisticated strategies. 
In random on/off control, we model the on/off control strategy as a Bernoulli trial, 
such that each BS continues to operate with probability $q$ and is turned off with probability $1-q$, 
independent of all the other base stations.

Instead of turning the BS off randomly, the BS may be turned off when the activity level is low, such as when the load or traffic demand is low. 
This study assumes that the BS's activity level is determined by the number of users in its coverage area. 
It should be noted that this strategy is premised on the assumption that all users have the same traffic requirements.

\ifdefined\singlecolumn
\begin{figure*}[htbp]
  \centering
  \begin{subfigure}[b]{0.49\linewidth}
  \vskip 0pt
  \includegraphics[width=\linewidth]{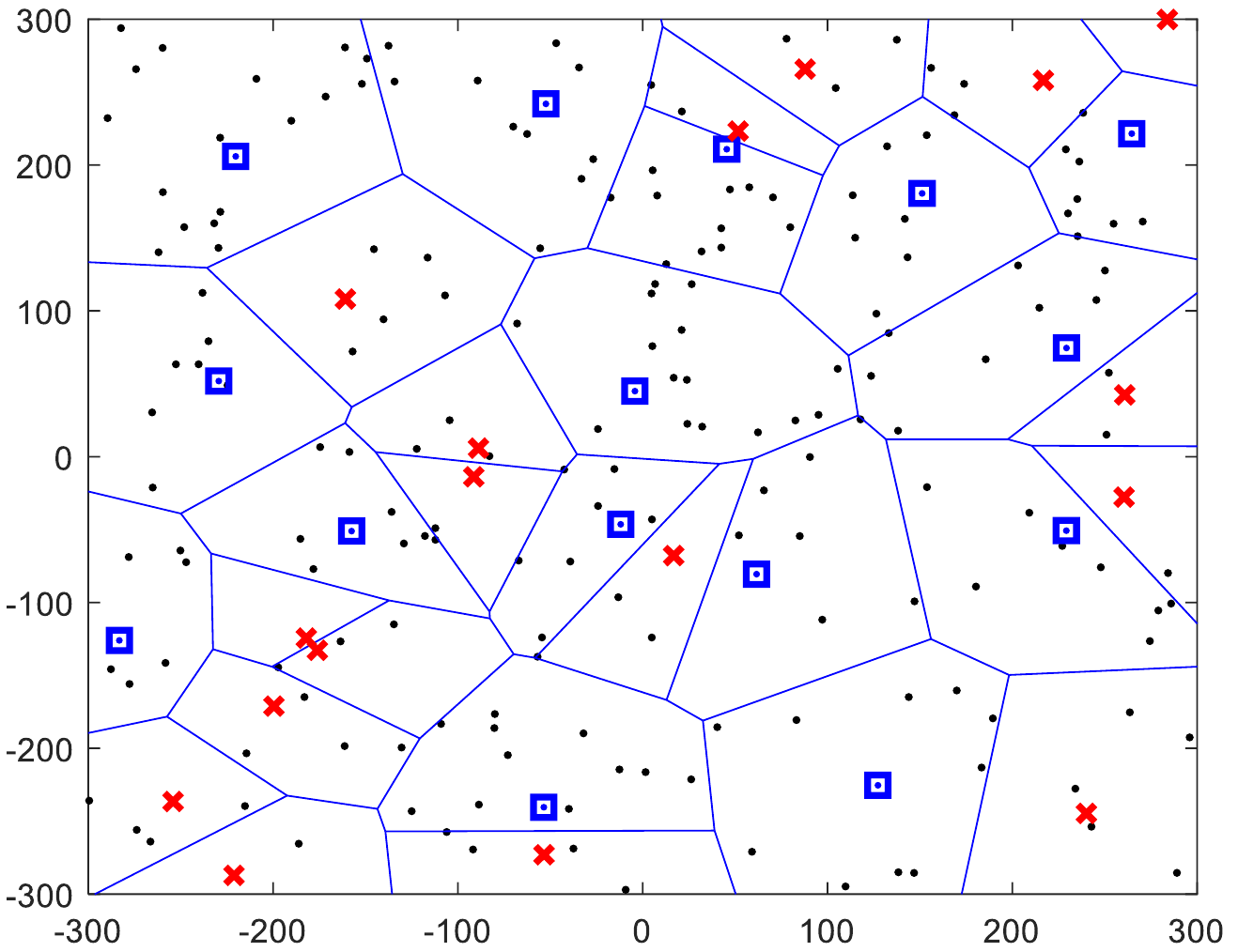}
  \vspace{3pt}
  \caption{Original cell boundaries\\ $\quad$ \\ $\quad$}
  \label{fig:BS_MS_distribution}
  \end{subfigure}
  \hfill
  \begin{subfigure}[b]{0.49\linewidth}
   \includegraphics[width=\linewidth]{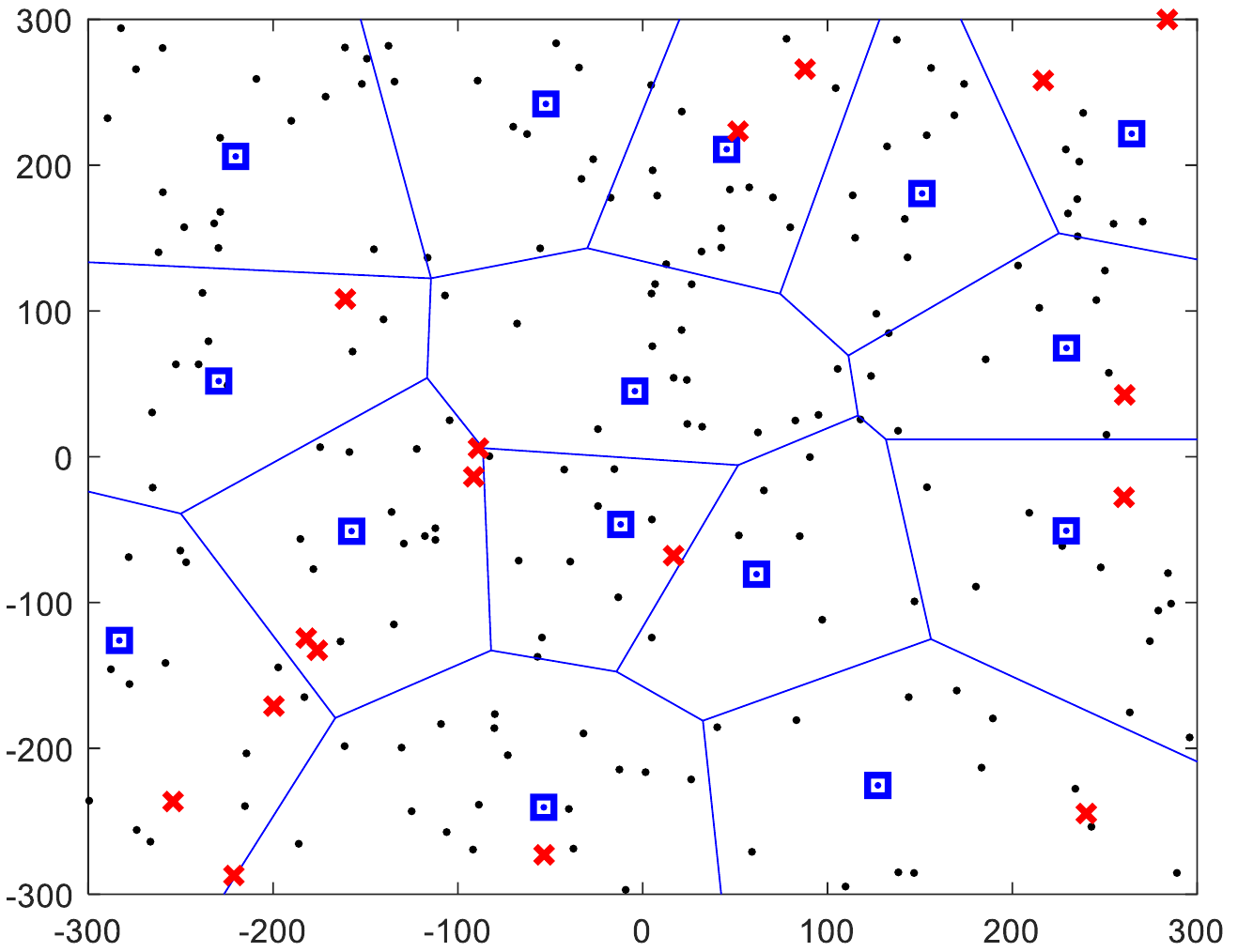}
  \vspace{3pt}
  \caption{Redrawn cell boundaries with user-number threshold-based on/off control where the user-number threshold $\theta$ is 5}
  \label{fig:BS_MS_distribution_w_onoff} 
  \end{subfigure}
  \caption{Poisson distributed BSs and users. Each user is associated with the nearest BS, where densities of BS and user are $\lambda_b=\num{1e-4}$ and $\lambda_u=\num{5e-4}$, respectively. The black dots are mobile users; the blue squares are the BS where the active number of users is above $\theta$; the red `x's are the BS where the active number of users is less than equal to $\theta$. The cell boundaries are shown and form a Voronoi tessellation.}
  \label{fig:voronoi_example}
  \vspace{-9pt}
\end{figure*}
\else
\fi

Fig.~\ref{fig:voronoi_example} shows the distribution of the BSs and users with cell boundaries. 
When deploying the user-number threshold-based on/off control scheme, BSs with fewer users than the threshold are deactivated. 
Users belonging to the BS of the off-mode are connected to their next-nearest BS, which is active (in the on-mode). 
Then, the cell boundaries are redrawn among the active BSs, as in Fig. \ref{fig:BS_MS_distribution_w_onoff}. 

Interestingly, 
when comparing the original and redrawn cell boundaries, 
it is identified that the cases where some BSs are very closely located together, 
which are the main weakness of the Poisson model~\cite{Jeffrey:2011}, are removed. 
The distribution of active BSs and redrawn cell boundaries become more similar to the real base station deployment shown in~\cite{Jeffrey:2011}. 
Therefore, the performance metrics of the network model with user-number threshold-based on/off control are expected to reflect the performance of actual network deployment more closely than a totally random network model or an ideal hexagonal grid model, assuming the same density of active BSs.

\subsection{Signal-to-Interference-plus-Noise Ratio}

We conduct an analysis on a typical mobile user located at the origin without loss of generality. 
The fading (power) between the typical user and the BS located at point $x$ is indicated by $h_x$, which is assumed to be drawn from an independent and identically distributed exponential function (Rayleigh fading). 
The standard path loss function is given by $l(x)=\|x\|^{\alpha}$, where $\alpha > 2$ is the path loss exponent. 
Hence, for the HetNet, the received power at a typical mobile user from a BS of the $i$-th tier located at point $x$ is $P_{t,i} h_x \|x\|^{-\alpha}$, where $h_x\backsim\exp(1)$ and $P_{t,i}$ is the transmit power of the BS in the $i$-th tier.
The resulting SINR expression assuming that the user connects to this BS is 
\be
\mbox{SINR}(x) = \frac{P_{t,i} h_x \|x\|^{-\alpha}}{\sum_{j=1}^{K}\sum_{y\in\Phi_j \setminus x} P_{t,j} h_y \|y \|^{-\alpha} +\sigma^2 },
\label{eq:sinr_expression}
\ee
where $\sigma^2$ is the constant additive noise power. 
Note that when $K = 1$, \eqref{eq:sinr_expression} becomes the SINR expression for a HomNet by omitting the subscripts $i$ and $j$. 

\subsection{Power Control}

A power control policy is adopted here to avoid creating coverage holes or areas where the SNR is below an acceptable level owing to switching off BSs. 
To ensure a similar coverage level as before deploying on/off control, we assume that all remaining active BSs transmit with power $P_t = \beta P_{T}$, where $\beta$ is a ratio that represents power control, and $P_{T}$ is the initial power prior to applying the on/off scheme. 

As the number of BSs being switched off increases, the cell size of the remaining BS increases; when the HomNet is thinned with the active probability of the on/off control $p_a$,
the cell size of the remaining BSs increases proportionally with $1/p_a$.
Along with the increased cell size, the distance to the cell-edge increases as well, with the rate of $1/\sqrt{p_a}$. 
To provide the equivalent service or SNR to the cell-edge users, even with the on/off scheme, the power control ratio $\beta$ must compensate for the increased distance with the rate of $1/\sqrt{p_a}$. 
Consequently, $\beta$ becomes $1/\sqrt{p_a^{\alpha}}$. 

For the HetNet, the same power control is applied with the average active probability of the whole network, that is, $p^{avr}_a$. 
Then, $\beta = 1/\sqrt{({p^{avr}_a})^{\alpha}}$ for all BSs of all tiers in the network. All active BSs of the $i$-th tier transmit with power $P_{t,i} = \beta P_{T,i}$, where $P_{T,i}$ is the initial transmit power of the $i$-th tier when the on/off scheme is disabled. 

\section{Homogeneous Network}
\label{sec:HomNet}

In this section, we investigate how user-number threshold-based on/off control affects the coverage probability of the HomNet. 
The coverage probability is defined as the probability that a typical user can achieve a target SINR, or the average fraction of users who at any time achieve the target SINR \cite{Jeffrey:2011}, which is expressed as
\be
P_{c} \left( T , \alpha, \lambda_b, \lambda_u, \theta \right) = \mathbb{P} \left[ \mbox{SINR} > T \right].
\ee
In this on/off control scheme,
when the typical BS has users in its coverage area fewer than or equal to the user-number threshold $\theta$, the BS is turned off. 
In this approach, BSs in densely populated areas will continue to provide services, and BSs in idle areas will be forced to rest, thereby reducing the overall impact of interference. 
Users belonging to BSs in the off-mode find another active BS that provides the next-highest link quality. 

Such users are likely to experience lower SINR and lower coverage probability despite the reduced interference, because more distant BSs serve such users. 
Then, the overall coverage probability can be improved by controlling $\theta$, such that the benefit of interference reduction for the entire network is greater than the adverse effect of deteriorating service quality of users in idle BSs.
To explore the effect of $\theta$ on the coverage probability, it is necessary to express the overall coverage probability as a function of $\theta$.

\subsection{Coverage Probability of Coordinated On/Off Control}

The network's overall coverage probability with user-number threshold-based on/off control can be obtained as follows: 
\ifdefined\singlecolumn
\be
P_{c} \left( T , \alpha, \lambda_b, \lambda_u, \theta \right) 
= \mathbb{E}_n \biggl[ P_{c} \left( T , \alpha, \lambda_b, \lambda_u, \theta | n \right) \biggr]
= \sum_{n=1}^{\infty} P_{c} \left( T , \alpha, \lambda_b, \lambda_u, \theta | n \right) 
\cdot p_n(\theta, \gamma),
\label{eq:overall_Pc}    
\ee
\else
\begin{multline}
P_{c} \left( T , \alpha, \lambda_b, \lambda_u, \theta \right) 
= \mathbb{E}_n \biggl[ P_{c} \left( T , \alpha, \lambda_b, \lambda_u, \theta | n \right) \biggr]  \\
= \sum_{n=1}^{\infty} P_{c} \left( T , \alpha, \lambda_b, \lambda_u, \theta | n \right) 
\cdot p_n(\theta, \gamma),
\label{eq:overall_Pc}    
\end{multline}
\fi
where $p_n(\theta,\gamma)$ is the probability that the user is connected to the $n$-th closest active BS for $\theta$ and $\gamma=\lambda_u/\lambda_b$. 
$P_{c} \left( T , \alpha, \lambda_b, \lambda_u, \theta | n \right)$ is the coverage probability conditioned on that the user is connected to the $n$-th closest active BS. 
The overall coverage probability is obtained by averaging $P_{c} \left( T , \alpha, \lambda_b, \lambda_u, \theta | n \right)$ with respect to $n$.

Prior to obtaining $p_n(\theta,\gamma)$, an investigation of some of the required metrics is presented. 
First, the probability that a typical BS contains $m$ users is given by \cite{Cao:2013,Mugume:2019}. 
It is derived from the cell size probability density function~(PDF) of the standard  Voronoi tessellation modeled by a gamma distribution~\cite{Pineda:2008}, and  
it is expressed as follows:
\ifdefined\singlecolumn
\be
\mathbb{P} (N=m | \gamma)  = \frac{\lambda_u^m (\mathcal{K}\lambda_b)^\mathcal{K} \Gamma(m+\mathcal{K})}{\Gamma(\mathcal{K}) m! (\lambda_u+K\lambda_b)^{m+\mathcal{K}}}   = \frac{\Gamma(m+\mathcal{K})}{\Gamma(\mathcal{K})m!}\left(1+\frac{\mathcal{K}}{\gamma}\right)^{-m} \left(1+\frac{\gamma}{\mathcal{K}} \right)^{-\mathcal{K}},
\label{eq:prob_num_users_in_BS}
\ee
\else
\begin{align}
\mathbb{P}& (N=m | \gamma)  \nonumber \\
&= \frac{\lambda_u^m (\mathcal{K}\lambda_b)^\mathcal{K} \Gamma(m+\mathcal{K})}{\Gamma(\mathcal{K}) m! (\lambda_u+K\lambda_b)^{m+\mathcal{K}}}  \nonumber \\
       &= \frac{\Gamma(m+\mathcal{K})}{\Gamma(\mathcal{K})m!}\left(1+\frac{\mathcal{K}}{\gamma}\right)^{-m} \left(1+\frac{\gamma}{\mathcal{K}} \right)^{-\mathcal{K}},
\label{eq:prob_num_users_in_BS}
\end{align}
\fi
where $N$ is the underlying random variable for the number of connected users in a typical BS,
$\Gamma(\mathcal{K}) = \int_0^{\infty}x^{\mathcal{K}-1}\mbox{exp}(-x)dx$ is the gamma function, and 
$\mathcal{K}~=~3.575$.

Then, when the user-number threshold-based on/off control scheme is deployed with $\theta$,
the active probability of a BS is expressed as follows:
\be
p_{a}\left(\theta, \gamma \right) = \mathbb{P}\left(N > \theta | \gamma \right) = 1- \mathbb{P}\left(N \leq \theta | \gamma \right).
\label{eq:active_prob}
\ee
The probability that a typical user connects to a BS other than the nearest BS, that is, $p_{{1}^\complement}(\theta,\gamma)$, is obtained next.
Note that $p_{{1}^\complement}(\theta,\gamma)$ is the complementary probability of $p_{{1}}(\theta,\gamma)$, which is the probability that the typical user is connected to the nearest active BS. 
$p_{{1}^\complement}(\theta,\gamma)$ is expressed as the ratio of the average total number of users located in the coverage area of the idle BSs to the average total number of users in the entire network, which is expressed as follows:
\begin{align}
p_{{1}^\complement}(\theta,\gamma) 
= \dfrac{\sum \limits_{m=1}^{\theta} m \mathbb{P}(N=m | \gamma) \lambda_b}{\lambda_u}
\label{eq:serve_BS_not_closest}
\end{align}
A typical user located in an idle BS finds the next-closest active BS. 
If all BSs close to the $(n-1)$-th order from the typical user are all idle, then the user connects to the $n$-th nearest active BS.
From \eqref{eq:active_prob} and \eqref{eq:serve_BS_not_closest}, the probability of a typical user connecting to the $n$-th closest active BS is expressed as
\ifdefined\singlecolumn
\be
p_n(\theta,\gamma)  = \left\{
        \begin{array}{ll}
          1-p_{{1}^\complement}(\theta,\gamma), & \hbox{for $n=1$;} \\
          p_{{1}^\complement}(\theta,\gamma) (1-p_{a}(\theta,\gamma))^{n-2} p_a(\theta,\gamma), & \hbox{for $n \geq 2$.}
        \end{array}
      \right.
      \label{eq:pn}
\ee
\else
\begin{multline}
p_n(\theta,\gamma)  = \\ \left\{
        \begin{array}{ll}
          1-p_{{1}^\complement}(\theta,\gamma), & \hbox{for $n=1$;} \\
          p_{{1}^\complement}(\theta,\gamma) (1-p_{a}(\theta,\gamma))^{n-2} p_a(\theta,\gamma), & \hbox{for $n \geq 2$.}
        \end{array}
      \right.
      \label{eq:pn}
\end{multline}
\fi


Now, we obtain the coverage probability given that the typical user connects to the $n$-th closest BS, that is, $P_{c} \left( T , \alpha, \lambda_b, \lambda_u, \theta | n \right)$.
To obtain the coverage probability,
we investigate the stochastic characteristics of
the distance from the typical user to the $n$-th closest BS, that is, $R_n$.

Note that the PDF of $R_n$ 
is not totally independent of the event that the $n$-th closest BS is active or idle in the user-number threshold-based on/off control scheme. 
For example, suppose a certain BS is active. 
In this case, the BS is likely to have a large coverage area containing more users than the threshold.
The average distance between the BS and its users is expected to be larger 
than that in the typical case. The greater the user-number threshold, 
the greater the average distance between the BS and its users.

Conversely, when a particular BS is idle, the BS is likely to have a small coverage area, and thus to have fewer users than the threshold.
Then, the distance between the idle BS and users in its coverage is likely to be smaller than in the typical case. 
Thus, for a sufficiently large $n$, 
when the BSs up to the $(n-1)$-th closest are all idle and the $n$-th closest BS is active, the distance to the $n$-th closest BS is likely to be less than the distance to the typical $n$-th closest BS.

However, it is demanding to obtain the exact expression of the PDF of $R_n$,  conditioned on not only the event that the $n$-th closest BS is active but also the event that the BSs up to the $(n-1)$-th closest order are inactive. 
Even if an exact expression is obtained, the expression is likely so complex that the final coverage probability is also very complicated. 
Therefore, it would be challenging to obtain a useful interpretation from the final results.
In this study, the PDF of $R_n$ under the conditions that the $n$-th closet BS is active and BSs up to the $(n-1)$-th closest order are inactive is approximated as the PDF of $R_n$ for the typical $n$-th closet BS.

The PDF of $R_n$ for the typical $n$-th closest BS can be derived using the cumulative distribution function~(CDF) of $R_n$. 
The complementary CDF (CCDF) of $R_n$ is equivalent to the probability that the circular area with radius $r$ contains fewer than $n$ users, which is expressed as follows :  
\be
\mathbb{P}(R_n \geq r) = \sum_{k=0}^{n-1} e^{-\lambda \pi r^2} \frac{(\lambda \pi r^2)^k}{k!}
\ee
Then, the CDF is $F_{R_n}(r) =  1 - \mathbb{P}(R_n \geq r) $ and
the PDF of $R_n$ is expressed as follows:
\be
f_{R_n}(r) = \frac{dF_r(r)}{dr} = e^{-\lambda_b \pi r^2}\frac{ 2 (\lambda_b \pi r^2 )^n} {r \Gamma(n)}.
\label{eq:pdf_of_dist_to_nth_BS}
\ee

We now state our first major result, the coverage probability of a typical user served by the $n$-th closest BS. 
\begin{theorem}
Conditioned on the event that the user is connected to the $n$-th closest BS, the probability that the user can achieve a target SINR $T$ is given by
\ifdefined\singlecolumn
\be
P_c(T,\alpha,\lambda_b,\lambda_u,\theta| n) = 
\frac{(\pi \lambda_b)^n}{\Gamma(n)} \int_0^{\infty}e^{-\pi\lambda_b v (1+ p_a( \theta, \gamma ) \rho(T,\alpha))-\frac{Tv^{\alpha/2}}{SNR}}v^{n-1} dv,
\ee
\else
\begin{multline}
P_c(T,\alpha,\lambda_b,\lambda_u,\theta| n) = \\
\frac{(\pi \lambda_b)^n}{\Gamma(n)} \int_0^{\infty}e^{-\pi\lambda_b v (1+ p_a( \theta, \gamma ) \rho(T,\alpha))-\frac{Tv^{\alpha/2}}{SNR}}v^{n-1} dv,
\end{multline}
\fi
where $SNR = P_t/\sigma^2$, and
\ifdefined\singlecolumn
$\rho(T,\alpha) = T^{2/\alpha} \int_{\infty}^{T^{-2/\alpha}} \frac{1}{1+u^{\alpha/2}} du$.
\else
\be
\rho(T,\alpha) = T^{2/\alpha} \int_{\infty}^{T^{-2/\alpha}} \frac{1}{1+u^{\alpha/2}} du.
\ee
\fi
\end{theorem}

\begin{IEEEproof}
The coverage probability is given by
\ifdefined\singlecolumn
\be
P_c(T,\alpha,\lambda_b,\lambda_u,\theta| n) = \mathbb{E}_r \left[ \mathbb{P} \left( \mbox{SINR} > T | r \right) \right]  = \int_{r>0} \mathbb{P}\left( \mbox{SINR} > T | r  \right) f_{R_n}(r) dr, 
\ee
\else
\begin{align}
P_c(T,\alpha,\lambda_b,\lambda_u,\theta| n) &= \mathbb{E}_r \left[ \mathbb{P} \left( \mbox{SINR} > T | r \right) \right]  \\
&= \int_{r>0} \mathbb{P}\left( \mbox{SINR} > T | r  \right) f_{R_n}(r) dr, \nonumber
\end{align}
\fi
where $f_{R_n}(r)$ is given by \eqref{eq:pdf_of_dist_to_nth_BS} and
the details of the exact derivation of $\mathbb{P}\left( \mbox{SINR} > T | r  \right)$
can be found in many previous studies that approach modeling wireless networks with stochastic geometry, such as \cite{Jeffrey:2011}, expressed as follows:
\ifdefined\singlecolumn
\ben
\mathbb{P}\left( \mbox{SINR} > T | r  \right) &=& \mathbb{P}\left( \left. \frac{h r^{-\alpha}}{\sigma^2 + I_r} > T  \right| r  \right)  
 = \exp \left( -\frac{T r^\alpha}{\mbox{SNR}}   \right)
\mathbb{E}_{I_{r}} \left[ \exp\left( -\frac{T r^\alpha}{P_t} I_r \right)  \right]  \nonumber \\
 &\stackrel{(a)}{=}& \exp \left( -\frac{T r^\alpha}{\mbox{SNR}}   \right) \exp \left( - \pi r^2
p_a( \theta, \gamma ) \lambda_b \rho \left( T, \alpha \right)   \right), 
\label{eq:prob_cov}
\een
\else
\begin{align}
&\mathbb{P}\left( \mbox{SINR} > T | r  \right) = \mathbb{P}\left( \left. \frac{h r^{-\alpha}}{\sigma^2 + I_r} > T  \right| r  \right)  \nonumber \\
&\qquad = \exp \left( -\frac{T r^\alpha}{\mbox{SNR}}   \right)
\mathbb{E}_{I_{r}} \left[ \exp\left( -\frac{T r^\alpha}{P_t} I_r \right)  \right]  \nonumber \\
&\qquad \stackrel{(a)}{=} \exp \left( -\frac{T r^\alpha}{\mbox{SNR}}   \right) \exp \left( - \pi r^2
p_a( \theta, \gamma ) \lambda_b \rho \left( T, \alpha \right)   \right), 
\label{eq:prob_cov}
\end{align}
\fi
where (a) follows from the Laplace transform of $I_r$ which is the thinned interference power with $p_a( \theta, \gamma ) \lambda_b$.
\end{IEEEproof}

\begin{corollary}
When $\alpha =4$, Theorem 1 can be simplified into the following form:
\ifdefined\singlecolumn
\be
P_c\left(T,4,\lambda_b,\lambda_u,\theta| n \right)
=\left( \frac{2T}{\mbox{SNR}}\right)^{n/2} (n-1)! 
\cdot \exp\left({\left(\pi\lambda_b \kappa\left(T\right)\right)^2}\over{{8T}\over{\rm SNR}}\right)
D_{-n} \left( {\pi\lambda_b \kappa\left(T\right)}\over{\sqrt{2T\over{\rm SNR}}} \right),
\label{eq:Pc_alpha_is_4}
\ee
\else
\begin{multline}
P_c\left(T,4,\lambda_b,\lambda_u,\theta| n \right)
=\left( \frac{2T}{\mbox{SNR}}\right)^{n/2} (n-1)!  \\
\cdot \exp\left({\left(\pi\lambda_b \kappa\left(T\right)\right)^2}\over{{8T}\over{\rm SNR}}\right)
D_{-n} \left( {\pi\lambda_b \kappa\left(T\right)}\over{\sqrt{2T\over{\rm SNR}}} \right),
\label{eq:Pc_alpha_is_4}
\end{multline}
\fi
where $D_{v}(z)$ is the parabolic cylinder function, and
\ifdefined\singlecolumn
\be
 \kappa (T) = 1+p_a(\theta, \gamma) \rho (T,4) = 1+p_a(\theta, \gamma) \sqrt{T}( \pi/ 2 -\tan^{-1}(1/\sqrt{T})).
\ee
\else
\begin{align}
 \kappa (T) &= 1+p_a(\theta, \gamma) \rho (T,4) \\ &= 1+p_a(\theta, \gamma) \sqrt{T}( \pi/ 2 -\tan^{-1}(1/\sqrt{T})). \nonumber
\end{align}
\fi
\end{corollary}
\begin{IEEEproof}
Theorem 1 can be evaluated according to (A6) of the integral table \cite{int_table}, expressed as
\ifdefined\singlecolumn
\be
\int_{0}^{\infty} x^n e^{-a^2 x^2 + b x} 
= (2a^2)^{(n+1)/2} n! \exp\left( \frac{b^2}{8 a^2}  \right) D_{-(n+1)}\left( -\frac{b}{a\sqrt{2}}  \right).
\ee
\else
\begin{multline}
\int_{0}^{\infty} x^n e^{-a^2 x^2 + b x} \nonumber \\
= (2a^2)^{(n+1)/2} n! \exp\left( \frac{b^2}{8 a^2}  \right) D_{-(n+1)}\left( -\frac{b}{a\sqrt{2}}  \right).
\end{multline}
\fi
Setting $a~=~\sqrt{T/SNR}$ and $b=-\pi\lambda_b(1+ p_a(\theta, \gamma) \rho(T,\alpha))$ yields \eqref{eq:Pc_alpha_is_4}.
\end{IEEEproof}

Note that $D_{v}(z)$ can be evaluated easily by modern calculators and software programs.
When $n=1$, that is, the user connects the closest BS,
\eqref{eq:Pc_alpha_is_4} yields the same result as \cite{Jeffrey:2011} with the special case $D_{-1}(z) = \exp(z^2/4) \sqrt{2 \pi} Q(z)$, where $Q(z)$ is the standard Gaussian tail probability.

The following corollary gives the more simplified form of the coverage probability of  Theorem 1 for the interference-limited case.
Thus, we can capture how the coverage probability is associated with the distance order of the serving BS. 
\begin{corollary}
If $\sigma^2 \rightarrow 0$ (or transmit power is increased sufficiently), the coverage probability of Theorem 1 can be simplified as
\be
P_{c} \left( T, \alpha, \gamma , \theta | n \right) = \frac{1}{\left( 1 + p_a(\theta, \gamma ) \rho\left(T,\alpha\right)\right)^n}.
\label{eq:Pc_inf_limited}
\ee
\end{corollary}
\begin{IEEEproof}
When $\sigma^2 \rightarrow 0$, through the
pre-known integral result $\int_{0}^{\infty} e^{-ax} x^{n-1} {\rm d}x = \Gamma(n)a^{-n}$,
Theorem 1 can be evaluated into the form of \eqref{eq:Pc_inf_limited}.
\end{IEEEproof}
From \eqref{eq:Pc_inf_limited}, it is interesting to note that the coverage probability decreases geometrically with respect to $n$. 
Note that \eqref{eq:Pc_inf_limited} does not depend on either $\lambda_b$ or $\lambda_u$, 
but it depends on their ratio, $\gamma$. 
Therefore, for the interference-limited case, $\lambda_b$ and $\lambda_u$ are substituted with $\gamma$ in the input arguments of the coverage probability.

Finally, from \eqref{eq:overall_Pc},
the overall coverage probability can be derived 
by averaging $P_{c} \left( T, \alpha, \lambda_b, \lambda_u, \theta | n \right)$ over $n$ with $p_n(\theta, \gamma)$.
For the general case of Theorem 1 and the case $\alpha = 4$ of Corollary 1, 
however, it is challenging to evaluate the overall coverage probability for $n$ from zero to infinity, owing to the complexity of the expression of $P_{c} \left( T, \alpha, \lambda_b, \lambda_u , \theta | n \right)$.
Therefore, 
to obtain a more simplified result, we assume the interference-limited case only, neglecting thermal noise.
It should be noted that thermal noise is not an important consideration in most modern cellular networks.
Thermal noise can be ignored inside the cell because it is very low compared to the desired signal output. At the edges of the cells, the interference is generally much greater, such that thermal noise can be neglected as well.

\begin{theorem}
For the interference-limited case, \eqref{eq:overall_Pc} is greatly simplified with $\rho\left(T,\alpha\right)$, $p_a(\theta, \gamma)$, and $p_1(\theta, \gamma)$, expressed as follows:
\begin{align}
P_{c} \left( T, \gamma, \theta, \alpha \right)
= \frac{1+ p_1(\theta, \gamma) \rho\left(T,\alpha\right)}{\left( 1+ \rho\left(T,\alpha\right)  \right)\left( 1+ p_a(\theta, \gamma) \rho\left(T,\alpha\right)  \right)}.
\label{eq:overall_Pc_inf_limited}
\end{align}
\label{the:overall_CP_HomNet}
\end{theorem}
\begin{IEEEproof}
From \eqref{eq:overall_Pc}, \eqref{eq:pn}, and \eqref{eq:Pc_inf_limited},
\ifdefined\singlecolumn
\begin{align}
&P_{c} \left( T, \gamma, \theta, \alpha \right)
= \sum_{n=1}^{\infty} \frac{1}{\left( 1 + p_a(\theta, \gamma) \rho\left(T,\alpha\right)\right)^n} p_n(\theta, \gamma) \nonumber \\
&= \frac{p_1(\theta, \gamma)}{1 + p_a(\theta, \gamma) \rho\left(T,\alpha\right)}  + \sum_{n=2}^{\infty} \frac{\left(1-p_1(\theta, \gamma)\right)\left(1-p_a(\theta, \gamma)\right)^{n-2}p_a(\theta, \gamma)}{\left( 1 + p_a(\theta, \gamma) \rho\left(T,\alpha\right)\right)^n} \nonumber \\
&= \frac{p_1(\theta, \gamma)}{1 + p_a(\theta, \gamma) \rho\left(T,\alpha\right)} + \frac{\left(1-p_1(\theta, \gamma)\right)p_a(\theta, \gamma)}{\left( 1 + p_a(\theta, \gamma) \rho\left(T,\alpha\right)\right)^2}
\sum_{n=2}^{\infty} \left( \frac{1-p_a(\theta, \gamma)}{1 + p_a(\theta, \gamma) \rho\left(T,\alpha\right)} \right)^{n-2} \nonumber \\
&= \frac{1}{1 + p_a(\theta, \gamma) \rho\left(T,\alpha\right)}  + \frac{\left(1-p_1(\theta, \gamma)\right)p_a(\theta, \gamma)}{\left( 1 + p_a(\theta, \gamma) \rho\left(T,\alpha\right)\right)^2} \frac{1}{1-\dfrac{1-p_a(\theta, \gamma)}{1 + p_a(\theta, \gamma). \rho\left(T,\alpha\right)}}. 
\label{eq:proof_of_overall_Pc}
\end{align}
\else
\begin{align}
& P_{c} \left( T, \gamma, \theta, \alpha \right)
= \sum_{n=1}^{\infty} \frac{1}{\left( 1 + p_a(\theta, \gamma) \rho\left(T,\alpha\right)\right)^n} p_n(\theta, \gamma) \nonumber \\
&= \frac{p_1(\theta, \gamma)}{1 + p_a(\theta, \gamma) \rho\left(T,\alpha\right)} \nonumber \\
&\qquad + \sum_{n=2}^{\infty} \frac{\left(1-p_1(\theta, \gamma)\right)\left(1-p_a(\theta, \gamma)\right)^{n-2}p_a(\theta, \gamma)}{\left( 1 + p_a(\theta, \gamma) \rho\left(T,\alpha\right)\right)^n} \nonumber \\
&= \frac{p_1(\theta, \gamma)}{1 + p_a(\theta, \gamma) \rho\left(T,\alpha\right)} \nonumber \\
& \hspace{5pt} + \frac{\left(1-p_1(\theta, \gamma)\right)p_a(\theta, \gamma)}{\left( 1 + p_a(\theta, \gamma) \rho\left(T,\alpha\right)\right)^2}
\sum_{n=2}^{\infty} \left( \frac{1-p_a(\theta, \gamma)}{1 + p_a(\theta, \gamma) \rho\left(T,\alpha\right)} \right)^{n-2} \nonumber \\
&= \frac{1}{1 + p_a(\theta, \gamma) \rho\left(T,\alpha\right)} \nonumber \\
& \hspace{5pt} + \frac{\left(1-p_1(\theta, \gamma)\right)p_a(\theta, \gamma)}{\left( 1 + p_a(\theta, \gamma) \rho\left(T,\alpha\right)\right)^2} \frac{1}{1-\dfrac{1-p_a(\theta, \gamma)}{1 + p_a(\theta, \gamma). \rho\left(T,\alpha\right)}}. 
\label{eq:proof_of_overall_Pc}
\end{align}
\fi
Expanding and rearranging \eqref{eq:proof_of_overall_Pc} gives \eqref{eq:overall_Pc_inf_limited}.
\end{IEEEproof}

Although we obtained Theorem {\ref{the:overall_CP_HomNet}} for user-number threshold-based on/off control, it can be used to evaluate another on/off control by changing only
$p_a(\theta, \gamma)$ and $p_1(\theta, \gamma)$
to the probabilities for the corresponding on/off control. 
The simplest example is random on/off control, which provides the baseline performance to compare to that of a more sophisticated strategy. 
When the random on/off control is deployed with the operating probability of $q$ (on), the network is simply modeled as a marked PPP with intensity $q\lambda_b$.

In addition, the coverage probability of random on/off control is essentially equivalent to the user-number threshold-based on/off control with $\theta = 0$, when the available BS density is independently thinned with~$q$. 
Then, the typical user always connects to the closest BS in the thinned network, such that $p_1(0, \gamma /q) = 1$. 
However, there is still a finite probability that a certain BS has no associated users in its coverage area. Such BSs are also turned off. 
The probability of the remaining active BS is 
$p_a(0,\gamma /q)$, and the resulting coverage probability becomes
\be
P_c(T,\gamma,q,\alpha) = \frac{1}{1+ p_a( 0, \gamma/q ) \rho(T,\alpha)}.
\label{eq:cov_prob_random}
\ee

Theorem {\ref{the:overall_CP_HomNet}} also applies to the evaluation of other on/off controls scheme with different activity level models of BSs, other than the number of users in a BS. 
This is done by only substituting 
$p_a(\theta, \gamma)$ and $p_1(\theta, \gamma)$ with the 
corresponding probabilities of the different arguments for the newly defined BS's activity model and the on/off control.  

\subsection{Optimal User-Number Threshold}
\label{sec:opt_threshold_Honet}

In this subsection, we analytically obtain the optimal user-number threshold  to maximize the overall coverage probability.
We consider only the interference-limited cases, as in Theorem 2.
The optimization problem is given as follows:
\be
\max_{\theta } P_{c} \left( T, \gamma, \theta, \alpha \right).
\ee
where $P_{c} \left( T, \gamma, \theta, \alpha \right)$ is the result of Theorem 2.

The easiest way to find a local maximum or minimum is differentiation.
However, the exact expressions of $p_1(\theta, \gamma)$ and $p_a(\theta, \gamma)$ are
so complicated that obtaining the derivative of $P_{c} \left( T, \gamma, \theta, \alpha \right)$ is challenging.
Thus, we use the approximation of $p_1(\theta, \gamma)$ and $p_a(\theta, \gamma)$.
Assuming that the size of a typical BS's coverage area is the average cell size of BSs in the unit area, that is, $1/\lambda_b$,
the probability of a typical BS containing $m$ users is approximated by the probability that there are $m$ users in the area of $1/\lambda_b$, expressed as
\be
\mathbb{P}(N=m | \gamma )  \approx  \tilde{\mathbb{P}}(N=m | \gamma ) = e^{-\gamma} \frac{\gamma^m}{m!},
\ee
which is the Poisson distribution with mean $\gamma$.
Then, from \eqref{eq:active_prob}, the approximation of  $p_a(\theta, \gamma)$  becomes the CCDF of the Poisson distribution as
$\tilde{p}_a(\theta, \gamma) = 1-\tilde{\mathbb{P}}(N \leq \theta | \gamma  )$.
In addition, $p_1(\theta, \gamma)$ is approximated as the CCDF of the Poisson distribution of $\gamma$ and $\theta$, expressed as
\ifdefined\singlecolumn
\begin{align}
\tilde{p}_{1}(\theta, \gamma)&= 1-{{\sum_{m=1}^{\theta} m \tilde{P}(N=m | \gamma ) \lambda_b}\over{\lambda_u}} 
= 1-\sum_{m=1}^{\theta} e^{-\gamma} \frac{\gamma^{m-1}}{(m-1)!} = 1-\sum_{m=0}^{\theta-1} e^{-\gamma}\frac{\gamma^m}{m!} \nonumber \\
& = 1-\tilde{\mathbb{P}}(N \leq \theta-1 | \gamma). 
\end{align}
\else
\begin{align}
\tilde{p}_{1}(\theta, \gamma)&= 1-{{\sum_{m=1}^{\theta} m \tilde{P}(N=m | \gamma ) \lambda_b}\over{\lambda_u}} \\
& = 1-\sum_{m=1}^{\theta} e^{-\gamma} \frac{\gamma^{m-1}}{(m-1)!} = 1-\sum_{m=0}^{\theta-1} e^{-\gamma}\frac{\gamma^m}{m!} \nonumber \\
& = 1-\tilde{\mathbb{P}}(N \leq \theta-1 | \gamma). \nonumber
\end{align}
\fi

Moreover,
owing to the complexity arising from the integer constraint on $\theta$,
obtaining the closed-form of derivatives of $\tilde{p}_a(\theta, \gamma)$ and $\tilde{p}_{1}(\theta, \gamma)$ with respect to $\theta$ is challenging.
Thus, $\tilde{p}_a(\theta, \gamma)$ and $\tilde{p}_{1}(\theta, \gamma)$ are approximated with a continuous distribution.
For sufficiently large $\gamma$, 
the Normal distribution~$\mathcal{N}(\mu = \gamma, \sigma^2 = \gamma)$ is an excellent approximation of the Poisson distribution of mean $\gamma$ 
if an appropriate continuity correction is performed.

With the approximation to the normal distribution,
\ifdefined\singlecolumn
\ben
\tilde{p}_a(\theta, \gamma) &=& 1-\tilde{\mathbb{P}}(N \leq \theta | \gamma  ) 
\approx 1 - F\left(\theta - \left. \frac{1}{2}  \right| \mu= \gamma, \sigma^2= \gamma \right) \mbox{ and } \nonumber \\
\tilde{p}_1(\theta, \gamma) &=& 1-\tilde{\mathbb{P}}(N \leq \theta-1 | \gamma) 
\approx 1 - F\left(\theta - \left. \frac{1}{2}  \right| \mu= \gamma, \sigma^2= \gamma \right),
\een
\else
\ben
\tilde{p}_a(\theta, \gamma) &=& 1-\tilde{\mathbb{P}}(N \leq \theta | \gamma  ) \nonumber \\
&\approx& 1 - F\left(\theta - \left. \frac{1}{2}  \right| \mu= \gamma, \sigma^2= \gamma \right) \mbox{ and } \nonumber \\
\tilde{p}_1(\theta, \gamma) &=& 1-\tilde{\mathbb{P}}(N \leq \theta-1 | \gamma) \nonumber \\
&\approx& 1 - F\left(\theta - \left. \frac{1}{2}  \right| \mu= \gamma, \sigma^2= \gamma \right),
\een
\fi
where $F(\cdot)$ is the normal cumulative distribution function.
Then, taking the derivative of $\tilde{p}_1(\theta, \gamma)$ and $\tilde{p}_a(\theta, \gamma)$ with respect to $\theta$ yields
\ifdefined\singlecolumn
\ben
\frac{d \tilde{p}_a(\theta, \gamma)}{ d\theta}
\approx - \frac{1}{\sqrt{2 \pi \gamma}} e^{ \left( -  \frac{(\theta-\gamma)^2+1/4}{2\gamma} \right)} e^{-\frac{\theta-\gamma}{2\gamma}} \mbox{   and } 
\frac{d \tilde{p}_1(\theta, \gamma)}{ d\theta}
\approx - \frac{1}{\sqrt{2 \pi \gamma}} e^{ \left( - \frac{(\theta-\gamma)^2+1/4}{2\gamma} \right)} e^{\frac{\theta-\gamma}{2\gamma}}.
\een
\else
\ben
\frac{d \tilde{p}_a(\theta, \gamma)}{ d\theta}
&\approx& - \frac{1}{\sqrt{2 \pi \gamma}} e^{ \left( -  \frac{(\theta-\gamma)^2+1/4}{2\gamma} \right)} e^{-\frac{\theta-\gamma}{2\gamma}} \mbox{   and } \nonumber \\
\frac{d \tilde{p}_1(\theta, \gamma)}{ d\theta}
&\approx& - \frac{1}{\sqrt{2 \pi \gamma}} e^{ \left( - \frac{(\theta-\gamma)^2+1/4}{2\gamma} \right)} e^{\frac{\theta-\gamma}{2\gamma}}.
\een
\fi
Using the above results, the derivative of the coverage probability with respect to $\theta$ is approximated as
\ifdefined\singlecolumn
\begin{multline}
\label{eq:derivative_of_Pc}
\frac{dP_{c}\left( T, \gamma, \theta, \alpha \right)}{d\theta} 
\approx -\frac{e^{ \left( -  \frac{(\theta-\gamma)^2+1/4}{2\gamma} \right)}}{\left(1+p_a(\theta, \gamma) \rho(T,\alpha)\right)^2 \sqrt{2 \pi \gamma}} \\
 \times \frac{\rho(T,\alpha)}{1+\rho(T,\alpha)} \biggl( e^{\frac{\theta-\gamma}{2\gamma}} \bigm( 1+ \tilde{p}_a(\theta, \gamma) \rho(T,\alpha)\bigm)
- e^{-\frac{\theta-\gamma}{2\gamma}} \bigm( 1+\tilde{p}_1(\theta, \gamma) \rho(T,\alpha) \bigm) \biggr).
\end{multline}
\else
\begin{multline}
\label{eq:derivative_of_Pc}
\frac{dP_{c}\left( T, \gamma, \theta, \alpha \right)}{d\theta}
\approx -\frac{e^{ \left( -  \frac{(\theta-\gamma)^2+1/4}{2\gamma} \right)}}{\left(1+p_a(\theta, \gamma) \rho(T,\alpha)\right)^2 \sqrt{2 \pi \gamma}} \\
 \times \frac{\rho(T,\alpha)}{1+\rho(T,\alpha)} \biggl( e^{\frac{\theta-\gamma}{2\gamma}} \bigm( 1+ \tilde{p}_a(\theta, \gamma) \rho(T,\alpha)\bigm)
\\
- e^{-\frac{\theta-\gamma}{2\gamma}} \bigm( 1+\tilde{p}_1(\theta, \gamma) \rho(T,\alpha) \bigm) \biggr).
\end{multline}
\fi
For $dP_{c}\left( T, \gamma, \theta, \alpha \right)/d\theta=0$,
\ifdefined\singlecolumn
\ben
\theta &=& \gamma \log \left( \frac{1+ \tilde{p}_1(\theta, \gamma) \rho(T,\alpha)}{1+\tilde{p}_a(\theta, \gamma) \rho(T,\alpha)} \right) + \gamma 
\stackrel{(a)}{=} \gamma \log \left( 1 + \frac{\tilde{\mathbb{P}}(N=\theta | \gamma)\rho(T,\alpha)}{1+\tilde{p}_a(\theta, \gamma) \rho(T,\alpha)} \right) + \gamma \\
& \stackrel{(b)}{\approx}& \gamma \log\left( 1 + \epsilon \right) + \gamma
\stackrel{(c)}{\approx} \gamma( 1 + \epsilon ), \nonumber
\een
\else
\ben
\theta &=& \gamma \log \left( \frac{1+ \tilde{p}_1(\theta, \gamma) \rho(T,\alpha)}{1+\tilde{p}_a(\theta, \gamma) \rho(T,\alpha)} \right) + \gamma \nonumber \\
&\stackrel{(a)}{=}& \gamma \log \left( 1 + \frac{\tilde{\mathbb{P}}(N=\theta | \gamma)\rho(T,\alpha)}{1+\tilde{p}_a(\theta, \gamma) \rho(T,\alpha)} \right) + \gamma \\
& \stackrel{(b)}{\approx}& \gamma \log\left( 1 + \epsilon \right) + \gamma
\stackrel{(c)}{\approx} \gamma( 1 + \epsilon ), \nonumber
\een
\fi
where (a) follows from 
\be
\tilde{p}_1(\theta, \gamma) =\tilde{p}_a(\theta,\gamma) + \tilde{\mathbb{P}}(N~=~\theta|\gamma),
\ee
(b) follows from $\tilde{\mathbb{P}}(N=\theta | \gamma)\rho(T,\alpha) \approx \epsilon$, where
$0< \epsilon \ll 1$ for sufficiently large $\gamma$, and 
(c) follows from $\log(1+\epsilon) \approx \epsilon$.

Considering the integer constraint on $\theta$, the optimal threshold $\theta_{opt}$ can be obtained
approximately by the closest integer to $\gamma$, as $\theta_{opt} = \lfloor \gamma + 0.5 \rfloor$ for a sufficiently large $\gamma$. 
It can be seen that \eqref{eq:derivative_of_Pc} is positive if $\theta \leq \theta_{opt}$, but negative otherwise.
In user-number threshold-based on/off control, deactivating BSs with 
users equal to or fewer than the average number of users per BS maximizes the overall coverage probability. 
Although users in the coverage area of low-activity BS (with fewer users than $\theta$) are heavily penalized, the benefits of decreased interference for the majority of the users outweigh the decreased performance for the minority. 

Furthermore, this result can provide an insight into the optimum on/off criteria for maximizing coverage probabilities in various on/off controls schemes. 
We have derived that the optimal user-number threshold is approximately the average number of users per BS. 
This is based on the assumption that all users have the same traffic requirements, and accordingly, the number of users in the BS represents the level of activity in BS. 
Thus, it can be assumed that the optimum on/off criterion for general coordinated On/Off control is near the average activity level of the BS. 



\section{Heterogeneous Network}

This section investigates a $K$-tier HetNet consisting of a multitier BS with different power and density. 
The analysis of the HetNet is much more complicated than that of the HomNet, because the network topologies of all tiers are closely intertwined.
For example, the cell size distribution in HomNets depends only on the BS density. However, in HetNets, the cell size distribution depends on, at least, the BS densities of every tier and their respective transmit powers.
Consequently, HetNet does not correspond to a standard Voronoi tessellation. Instead, it follows a multiplicatively weighted Voronoi diagram \cite{Dhillon:2012}, an extension of the standard Voronoi tessellation.

\subsection{Coverage Probability of Coordinated On/Off Control}

Owing to the highly complex topology of a HetNet,
it is very complicated to obtain the coverage probability.
In particular, applying coordinated BS on/off control for a HetNet makes this even more complicated. 
First, it is significantly challenging to obtain an accurate expression or model the probability that the typical BS of each tier contains $m$ users.  
Although the general expression of the corresponding probability for the HomNet has been found 
as in \eqref{eq:prob_num_users_in_BS}, it has not yet been found for the HetNet. 
This probability is used to obtain the active probability of a BS for each tier and the probability of a typical user connecting to the $n$-th closest BS for each tier in the HetNet.

To determine the probability that the typical BS of each tier contains $m$ users, each tier's cell size distribution should be provided.
\cite{Cao:2013} showed that the cell size distribution of each tier for the HetNet could be modeled as the gamma distribution as in the HomNet. 
However, a general expression has not yet been presented. 
Whenever the network topological parameters, such as densities or transmit powers, vary, we must conduct numerical fitting for every case and tier. 
Instead, this study suggests to consider each tier in a HetNet as a HomNet with a weighted density. 
Then, the probability that the typical BS of each tier contains $m$ users
can be accurately predicted by its corresponding expression of the HomNet case as in  \eqref{eq:prob_num_users_in_BS}. 

First, let us consider how the overall $K$-tier HetNet appears from the perspective of the typical user in the BS of the $i$-th tier. 
In the HetNet, owing to the differences in the transmit powers over the tiers, 
the user chooses its serving BS by comparing the weighted distance. For example, a user may choose a macro BS located far away, rather than the nearest micro BS. 
The typical user connects to the BS of the $i$-th tier with the distance $r_i$ if
\be
\frac{r_i}{w_i} < \frac{r_j}{w_j} \quad \forall j \neq i, 
\mbox{ where } w_i = P_{t,i}^{1/\alpha} \mbox{ and }  w_j = P_{t,j}^{1/\alpha}.
\ee
Thus, for a typical user connected to the BS of the $i$-th tier, the BS of the $j$-th tier with distance $r_j$ is equivalent to a BS of the $i$-th tier with the distance $r_i = r_j(w_i /w_j) $ in terms of the signal level \cite{Cao:2013}.
For this user, the $K$-tier HetNet is equivalent to a HomNet
with a density of
\be
\bar{\lambda}_{i} = \lambda_{i} + \sum_{j=1/i}^{K} \frac{ \lambda_{j} w_j^2}{w_i^2} =  \lambda_{i} \left( \frac{ \sum_{j=1}^{K} \lambda_j w_j^2 }{\lambda_{i} w_i^2} \right).
\label{eq:weighted_distance}
\ee
Then, by defining the probability that a typical user connects to the BS of the $i$-th tier as
\ben
q_i = \frac{\lambda_i}{\bar{\lambda}_i}= \frac{\lambda_i w_i^2}{\sum_j\lambda_j w_j^2},
\label{eq:prob_of_tier}
\een
\eqref{eq:weighted_distance} is simplified to $\bar{\lambda}_i = \lambda_i / q_i$. 

Consequently, the $K$-tier  HetNet is  equivalent  to  a  HomNet with a weighted density of $\bar{\lambda}_i$ from the perspective of the $i$-th tier. 
Accordingly, the probability that the typical BS of the $i$-th tier in a HetNet includes $m$ users can be estimated using $\mathbb{P} (N=m | \bar{\gamma}_i)$ of \eqref{eq:prob_num_users_in_BS} in the HomNet, where 
$\bar{\gamma}_i = \lambda_u/ \bar{\lambda}_i$. 
Then, for the $i$-th tier of the HetNet, the probability of a BS being active, that is, $p_{a,i}(\theta_i,\bar{\gamma}_i)$, and 
the probability that a typical user connects to a BS other than the nearest BS, that is, 
$p_{{1}^\complement,i}(\theta_i,\bar{\gamma}_i))$, follow the same forms as 
\eqref{eq:active_prob} and \eqref{eq:serve_BS_not_closest}, respectively.
For the $i$-th tier, we 
only substitute $\theta$ and $\gamma$ with $\theta_i$ and $\bar{\gamma}_i$. 
Note that $\theta_i$ is the user-number threshold of the $i$-th tier. It  is  assumed  that  the  BS  on/off  control  is  operated  independently  for  each  tier with  an  independent user-number threshold. 

Then, by averaging $p_{a,i}(\theta_i,\bar{\gamma}_i)$ and
$p_{{1}^\complement,i}(\theta_i,\bar{\gamma}_i))$ for all tiers, we obtain 
\ifdefined\singlecolumn
\be
p_a^{avr}(\{\theta \}, \{\bar{\gamma} \}) =\sum_{j=1}^{K} p_{a,j} (\theta_j, \bar{\gamma}_j ) q_j, \quad \mbox{ and } \quad 
p^{avr}_{1^{\complement}}(\{\theta\},\{\bar{\gamma} \}) = \sum_{j=1}^{K} p_{{1}^\complement,j}(\theta_j,\bar{\gamma}_j)) q_j,
\ee
\else
\ben
p_a^{avr}(\{\theta \}, \{\bar{\gamma} \}) &=&\sum_{j=1}^{K} p_{a,j} (\theta_j, \bar{\gamma}_j ) q_j, \quad \mbox{ and } \nonumber \\
p^{avr}_{1^{\complement}}(\{\theta\},\{\bar{\gamma} \}) &=& \sum_{j=1}^{K} p_{{1}^\complement,j}(\theta_j,\bar{\gamma}_j)) q_j,
\een
\fi
where $p^{avr}_a(\{\theta\},\{\bar{\gamma}\})$ is the probability of a BS being active over all tiers, and  $p^{avr}_{1^{\complement}}(\{\theta\},\{\bar{\gamma}\})$ is the probability that a typical user connects to a BS other than the nearest BS 
over all tiers for $\{\theta\}=\{\theta_1,\ldots \theta_K\}$ and $\{\bar{\gamma}\} = \{\bar{\gamma}_1,\ldots \bar{\gamma}_K \}$.
Moreover, we can represent the probability that a typical user connects to the $n$-th closest active BS as
\ifdefined\singlecolumn
\be
p_{n}(\{\theta\},\{\bar{\gamma}\})  = \left\{
        \begin{array}{lll}
          1-p^{avr}_{1^{\complement}}(\{\theta\},\{\bar{\gamma} \}), & \hbox{for $n=1$;} \\ \\
          p^{avr}_{1^{\complement}}(\{\theta\},\{\bar{\gamma} \})  (1-p^{avr}_a(\{\theta\},\{\bar{\gamma} \}))^{n-2} 
          p^{avr}_a(\{\theta\},\{\bar{\gamma} \}), & \hbox{for $n \geq 2$.}
        \end{array}
      \right.
      \label{eq:pn_HetNet}
\ee
\else
\begin{multline}
p_{n}(\{\theta\},\{\bar{\gamma}\})  = \\ \left\{
        \begin{array}{lll}
          1-p^{avr}_{1^{\complement}}(\{\theta\},\{\bar{\gamma} \}), & \hbox{for $n=1$;} \\ \\
          p^{avr}_{1^{\complement}}(\{\theta\},\{\bar{\gamma} \})  (1-p^{avr}_a(\{\theta\},\{\bar{\gamma} \}))^{n-2} \\ \qquad \qquad \times p^{avr}_a(\{\theta\},\{\bar{\gamma} \}), & \hbox{for $n \geq 2$.}
        \end{array}
      \right.
      \label{eq:pn_HetNet}
\end{multline}
\fi

In the following, we step through the same procedure of deriving the coverage probability in the HomNet, as in \eqref{eq:overall_Pc}, for the HetNet.
Despite the complexity of the analytical derivation, the results are quite simple and provide proof for the previous discussion. 
The results validate the previous discussion that the $i$-th  tier  in  the HetNet  can  be  analyzed  as  the HomNet with weighted density $\bar{\lambda}_i$. 
Finally, the analytical results show that the overall coverage probability expression of the HetNet has the form of a linear combination of HomNets.

Now, in order to obtain the coverage probability conditioned on that the user connects to the $n$-th closest BS in the HetNet,
the PDF of the distance to the $n$-th closest BS is obtained first, conditioned on the event that the $n$-th closest BS belongs to $i$-th tier.
\begin{lemma} \label{lem:pdf_of_dist_hetnet}
Conditioned on the event that the $n$-th closest BS,~$x_n$, belongs to the $i$-th tier, the PDF of  the distance to $x_n$, i.e. $R_{n|i}$, is
given by
\be
f_{R_{n|i}} (r) =  e^{\bar{\lambda}_i  \pi r^2 }  \frac{2\left( \bar{\lambda}_i \pi r^2 \right)^n}{r \Gamma(n)}.
\label{eq:pdf_of_dist_hetnet}
\ee
\label{lem:pdf_of_distance_HetNet}
\end{lemma}
\begin{IEEEproof}
See Appendix \ref{sec:lemma1}
\end{IEEEproof}

It is interesting to note that the PDF of $R_{n|i}$ in \eqref{eq:pdf_of_dist_hetnet} has the same form as \eqref{eq:pdf_of_dist_to_nth_BS}, the result for the HomNet.
Lemma~\ref{lem:pdf_of_distance_HetNet} also agrees with the discussion that the HetNet is equivalent to a HomNet with a density of $\bar{\lambda}_i$ from the perspective of the $i$-th tier.

Another necessary quantity is the Laplace transform of the interference power.
To obtain $\mathbb{P}\left( \mbox{SINR} > T | r  \right)$ as in \eqref{eq:prob_cov},
the Laplace transform of interference for the HetNet is required.
\begin{lemma} \label{lem:laplace_tf_jth_intf}
For a typical user connected to the BS of the $i$-th tier, the Laplace transform of interference is expressed as
\be
\mathbb{E}_{I_r} \left[  \exp \left( - \frac{Tr^\alpha}{P_{t,i}} I_r  \right) \right]
=  \exp \left( - \pi r^2 \bar{\lambda}_i  p^{avr}_a(\{\theta\},\{\bar{\gamma} \}) \rho(T,\alpha) \right).
\ee
\end{lemma}

\begin{IEEEproof}
See Appendix \ref{sec:lemma2}
\end{IEEEproof}
The result of Lemma \ref{lem:laplace_tf_jth_intf} also has the same form as \eqref{eq:prob_cov} for the HomNet, 
with only substituting $\lambda_j$ to $\bar{\lambda}_j$ and $p_a(\theta,\gamma)$ to $p^{avr}_a(\{\theta\},\{\bar{\gamma}\})$.
The interference power experienced by the typical user connected to the BS of the $i$-th tier
is equivalent to that of the HomNet with the density $\bar{\lambda}_i$, thinned with the average probability of being active $p^{avr}_a(\{\theta\},\{\bar{\gamma}\})$.

Lemmas 1 and 2 can now be used to obtain coverage probability under the assumption that the user is connected to the $n$-th closest BS. 
The following theorem presents the results:
\begin{theorem}
\label{th:Pc_HetNet}
In the HetNet, conditioned on the event that the user is connected to the $n$-th closest BS, the probability that the user can achieve a target SINR $T$ is given by
\ifdefined\singlecolumn
\be
P_c(T,\{\theta\},\{\bar{\lambda}\},\lambda_u,\alpha | n ) = 
\sum_{i=1}^K q_i \frac{(\pi \bar{\lambda}_i)^n}{\Gamma(n)} \int_0^{\infty}e^{-\pi \bar{\lambda}_i v \left(1+ {p}^{avr}_a\left(\{\theta\},\{\bar{\gamma}\}\right) \rho(T,\alpha)\right)-\frac{Tv^{\alpha/2}}{SNR}}v^{v-1} dv,
\label{eq:Pc_HetNet}
\ee
\else
\begin{multline}
P_c(T,\{\theta\},\{\bar{\lambda}\},\lambda_u,\alpha | n ) = \\
\sum_{i=1}^K q_i \frac{(\pi \bar{\lambda}_i)^n}{\Gamma(n)} \int_0^{\infty}e^{-\pi \bar{\lambda}_i v \left(1+ {p}^{avr}_a\left(\{\theta\},\{\bar{\gamma}\}\right) \rho(T,\alpha)\right)-\frac{Tv^{\alpha/2}}{SNR}}v^{v-1} dv,
\label{eq:Pc_HetNet}
\end{multline}
\fi
where $\{\bar{\lambda}\} = \{\bar{\lambda}_1 \ldots \bar{\lambda}_K\}$.
\end{theorem}

\begin{IEEEproof}
From Lemma \ref{lem:pdf_of_dist_hetnet} and Lemma \ref{lem:laplace_tf_jth_intf},
the coverage probability when a typical user connects to the $n$-th closest BS of $i$-th tier can be obtained  in the same form as
in Theorem 1, by substituting $\lambda$ with $\bar{\lambda}_i$ and $p_a(\theta,\gamma)$ with $p^{avr}_a(\{\theta\},\{\bar{\gamma}\})$.
Then, averaging this result over all tiers with $\mathbf{q}=\{q_1,\ldots,q_K\}$ yields \eqref{eq:Pc_HetNet}.
\end{IEEEproof}

Theorem 3 has an important implication on not only the analysis of coordinated on/off control for the HetNet, but also on the general stochastic geometry analysis of the HetNet.  From Theorem 3, it is identified that the coverage probability of the HetNet has the form of a linear combination of HomNets, where 
each HomNet corresponds to each tier of the HetNet with weighted density. 
This approach is also applicable to other performance indicators, such as average achievable rates and energy efficiency of HetNets, where they could be represented in the form of a linear combination of HomNets. 
Furthermore, for types of issues and applications other than BS on/off control, this approach also makes it easier to derive the numerical analysis of a HetNet based on the HomNet result.

When $\alpha=4$, Theorem \ref{th:Pc_HetNet} can be evaluated similarly to \eqref{eq:Pc_alpha_is_4} of the HomNet,
as follows:
\ifdefined\singlecolumn
\begin{multline}
P_{c} \left( T, \{\theta\}, \{\bar{\lambda}\},\lambda_u, 4 | n \right) \\
=\sum_{i=1}^{K} q_i \left( \frac{2T}{\mbox{SNR}}\right)^{n/2} (n-1)! 
\exp\left({\left(\pi \bar{\lambda}_i \kappa \left(T\right)\right)^2}\over{{8T}\over{\rm SNR}}\right)
D_{-n} \left( {\pi \bar{\lambda}_i \kappa\left(T\right)}\over{\sqrt{2T\over{\rm SNR}}} \right),
\label{eq:Pc_alpha_is_4_HetNet}
\end{multline}
\else
\begin{multline}
P_{c} \left( T, \{\theta\}, \{\bar{\lambda}\},\lambda_u, 4 | n \right)
=\sum_{i=1}^{K} q_i \left( \frac{2T}{\mbox{SNR}}\right)^{n/2} (n-1)!  \\
\cdot \exp\left({\left(\pi \bar{\lambda}_i \kappa \left(T\right)\right)^2}\over{{8T}\over{\rm SNR}}\right)
D_{-n} \left( {\pi \bar{\lambda}_i \kappa\left(T\right)}\over{\sqrt{2T\over{\rm SNR}}} \right),
\label{eq:Pc_alpha_is_4_HetNet}
\end{multline}
\fi
where
\be
 \kappa (T) = 1+p^{avr}_a (\{\theta\},\{\bar{\gamma}\}) \sqrt{T}( \pi/ 2 -\mbox{arctan}(1/\sqrt{T})).
\ee

Furthermore, for the interference-limited case with $\sigma^2 \rightarrow 0$,
Theorem \ref{th:Pc_HetNet} can be evaluated in the same form as \eqref{eq:Pc_inf_limited} as follows:
\be
P_{c} \left( T,\{\theta\}, \{\bar{\gamma}\}, \alpha, | n \right) = \frac{1}{\left( 1 + p^{avr}_a \left(\{\theta\},\{\bar{\gamma}\}\right) \rho\left(T,\alpha\right)\right)^n}.
\label{eq:Pc_inf_limited_Het}
\ee
Averaging this over $n$ follows the same procedure as the proof for Theorem 2, and its result also has the same form as \eqref{eq:overall_Pc_inf_limited}, as follows:
\ifdefined\singlecolumn
\be
P_{c} \left( T, \{\theta\}, \{\bar{\gamma}\}, \alpha \right) 
= \frac{1+ p^{avr}_1 (\{\theta\},\{\bar{\gamma}\}) \rho\left(T,\alpha\right)}{\left( 1+ \rho\left(T,\alpha\right)  \right)\left( 1+ p^{avr}_a (\{\theta\},\{\bar{\gamma} \})\rho\left(T,\alpha\right)  \right)}.
\label{eq:pc_hetnet}
\ee
\else
\begin{multline}
P_{c} \left( T, \{\theta\}, \{\bar{\gamma}\}, \alpha \right) \\
= \frac{1+ p^{avr}_1 (\{\theta\},\{\bar{\gamma}\}) \rho\left(T,\alpha\right)}{\left( 1+ \rho\left(T,\alpha\right)  \right)\left( 1+ p^{avr}_a (\{\theta\},\{\bar{\gamma} \})\rho\left(T,\alpha\right)  \right)}.
\label{eq:pc_hetnet}
\end{multline}
\fi

\subsection{Optimal User-Number Threshold}

As for the HomNet, we approximate $p_{1,i}(\theta_i,\bar{\gamma}_i)$ and $p_{a,i}(\theta_i,\bar{\gamma}_i)$ to obtain the optimal thresholds through the derivatives.
For the approximation, it is assumed that the size of the typical BS's coverage area of the $i$-th tier is equivalent to the average cell size of the BS of the $i$-th tier in the unit area, that is, $1/\bar{\lambda}_i$.
Then, the probability of a typical BS of the $i$-th tier contains $m$ users is approximated by the probability that
the number of users in an area of $1/\bar{\lambda}_i$, expressed as
\be
\tilde{\mathbb{P}}(N=m | \bar{\gamma}_i ) = e^{-\bar{\gamma}_i} \frac{\bar{\gamma}_i^m}{m!}.
\ee
Then, as in the same form of the HomNet,
$p_{a,i}(\theta_i,\bar{\gamma}_i)$ and $p_{1,i}(\theta_i,\bar{\gamma}_i)$ are approximated as 
\ifdefined\singlecolumn
\ben
\tilde{p}_{a,i}(\theta_i, \bar{\gamma}_i)
&=& 1-\tilde{\mathbb{P}}(N \leq \theta_i \bigm| \bar{\gamma}_i  ) 
\approx 1 - F\left(\theta_i + \left. \frac{1}{2}  \right| \mu= \bar{\gamma}_i, \sigma^2= \bar{\gamma}_i \right) \mbox{ and }  \nonumber \\
\tilde{p}_{1,i}(\theta_i, \bar{\gamma}_i) &=& 1-\tilde{\mathbb{P}}(N \leq \theta_i-1 | \bar{\gamma}_i) 
\approx 1 - F\left(\theta_i - \left. \frac{1}{2}  \right| \mu= \bar{\gamma}_i, \sigma^2= \bar{\gamma}_i \right), 
\een
\else
\ben
\tilde{p}_{a,i}(\theta_i, \bar{\gamma}_i)
&=& 1-\tilde{\mathbb{P}}(N \leq \theta_i \bigm| \bar{\gamma}_i  ) \nonumber \\
&\approx& 1 - F\left(\theta_i + \left. \frac{1}{2}  \right| \mu= \bar{\gamma}_i, \sigma^2= \bar{\gamma}_i \right) \mbox{ and } \nonumber \\
\tilde{p}_{1,i}(\theta_i, \bar{\gamma}_i) &=& 1-\tilde{\mathbb{P}}(N \leq \theta_i-1 | \bar{\gamma}_i). \nonumber \\
&\approx& 1 - F\left(\theta_i - \left. \frac{1}{2}  \right| \mu= \bar{\gamma}_i, \sigma^2= \bar{\gamma}_i \right), 
\een
\fi
respectively.
Also, the average of these approximated probabilities are denoted by $\tilde{p}^{avr}_a(\{\theta\},\{\bar{\gamma}\})$ and $\tilde{p}^{avr}_1(\{\theta\},\{\bar{\gamma}\})$.

Note that $\theta_i$ is only associated with $\tilde{p}_{a,i}(\theta_i, \bar{\gamma}_i)$
and $\tilde{p}_{1,i}(\theta_i, \bar{\gamma}_i)$, and is totally independent of
$\tilde{p}_{a,j}(\theta_j, \bar{\gamma}_j)$ and $\tilde{p}_{1,j}(\theta_j, \bar{\gamma}_j)$
for all $j \neq i$.
Thus, when taking the partial derivatives of  $\tilde{p}^{avr}_a(\{\theta\},\{\bar{\gamma}\})$ and $\tilde{p}^{avr}_1(\{\theta\},\{\bar{\gamma}\})$ with respect to $\theta_i$,
every term for the $j$-th tier ($j \neq i$) becomes zero, and only the derivatives of $\tilde{p}_{a,i}(\theta_i, \bar{\gamma}_i)$ and $\tilde{p}_{1,i}(\theta_i, \bar{\gamma}_i)$ are left, respectively.

Consequently, obtaining the optimal $\theta_i$ follows the same procedure in the HomNet analysis.
For $dP_{c}\left( T, \{\theta\},\{\bar{\gamma}\}, \alpha \right)/d\theta_i=0$, 
\ifdefined\singlecolumn
\ben
\theta_i &=& \bar{\gamma}_i \log \left( \frac{1+ p^{avr}_1(\{\theta\},\{\gamma\}) \rho(T,\alpha)}{1+p^{avr}_a(\{\theta\},\{\gamma\}) \rho(T,\alpha)} \right) + \bar{\gamma}_i \nonumber \\
&\stackrel{(a)}{=}& \bar{\gamma}_i \log \left( 1 + \frac{\sum_{j=1}^K \tilde{\mathbb{P}}(N=\theta_j | \bar{\gamma}_j)q_j\rho(T,\alpha)}{1+p^{avr}_a(\{\theta\},\{\bar{\gamma}\}) \rho(T,\alpha)} \right) + \bar{\gamma}_i \nonumber \\
& \stackrel{(b)}{\approx}& \bar{\gamma}_i \log\left( 1 + \epsilon \right) + \bar{\gamma}_i
\stackrel{(c)}{\approx} \bar{\gamma}_i( 1 + \epsilon ),
\een
\else
\ben
\theta_i &=& \bar{\gamma}_i \log \left( \frac{1+ p^{avr}_1(\{\theta\},\{\gamma\}) \rho(T,\alpha)}{1+p^{avr}_a(\{\theta\},\{\gamma\}) \rho(T,\alpha)} \right) + \bar{\gamma}_i \nonumber \\
&\stackrel{(a)}{=}& \bar{\gamma}_i \log \left( 1 + \frac{\sum_{j=1}^K \tilde{\mathbb{P}}(N=\theta_j | \bar{\gamma}_j)q_j\rho(T,\alpha)}{1+p^{avr}_a(\{\theta\},\{\bar{\gamma}\}) \rho(T,\alpha)} \right) + \bar{\gamma}_i \nonumber \\
& \stackrel{(b)}{\approx}& \bar{\gamma}_i \log\left( 1 + \epsilon \right) + \bar{\gamma}_i
\stackrel{(c)}{\approx} \bar{\gamma}_i( 1 + \epsilon ),
\een
\fi
where (a) follows from
\be
p^{avr}_1(\{\theta\},\{\bar{\gamma}\})
= p^{avr}_a(\{\theta\},\{\bar{\gamma}\}) + 
\sum_{j=1}^{K}\tilde{\mathbb{P}}(N~=~\theta_j|\bar{\gamma}_j) q_j
\ee
(b) follows from $\sum_{j=1}^K\tilde{\mathbb{P}}(N=\theta | \bar{\gamma}_j)q_j\rho(T,\alpha) \approx \epsilon$, 
where
$0< \epsilon \ll 1$ for sufficiently large $\bar{\gamma}_j$~$(\forall~j)$, and
(c) follows from $\log(1+\epsilon) \approx \epsilon$.
As in Section \ref{sec:opt_threshold_Honet}, 
the optimal threshold $\theta_{i,opt}$ can be obtained approximately as the closest integer to $\bar{\gamma}_i$, as $\theta_{i,opt} = \lfloor \bar{\gamma}_i + 0.5 \rfloor$, for  a  sufficiently  large $\bar{\gamma}_i$.

\section{Numerical Results}
\label{sec:numerical result}

In this section, we present the numerical results of both the analytical results and the Monte Carlo simulation results.
For the simulation of the HomNet, the locations of the BSs and users are distributed according to a PPP in a $6000~m \times 6000~m$ grid, with 1000 trials. 
It is assumed that $\lambda_b = \num{1e-4}$, $\lambda_u =  \num{1e-3}$, $\gamma =  10$, $T=0$~dB, $\alpha=4$, and $\theta$ varies from 0 to 19.

The initial transmit power for the BS is set according to the SNR of a cell-edge user $SNR_{e}$ \cite{Dhillon:2012}. 
It is expressed as $P_T = SNR_{e} d_{e}^{\alpha}\sigma^2$, where
$d_{e}$ is the cell-edge user distance. 
We define a cell-edge user as a user whose distance to the serving BS is at the 5th percentile of the CDF of user distance. 
Then, $\mathbb{P}(D \geq d_{e}) = 0.05$, where $D$ is the underlying random variable. 
For BSs distributed with a PPP of $\lambda_b$, $\mathbb{P}(D \geq d_{e}) = \exp(\lambda_b \pi d_{e}^2)$ giving $d_{e} = \sqrt{-\log(0.05) / ( \pi \lambda_b )}$. Assuming the interference-limited case, $SNR_{e}$ is assumed to be 10 dB more than $T$. The noise is assumed to have unit power for simplicity, as $\sigma^2 = 1$. 

\begin{figure}[!htb]
\centering
  \includegraphics[width=\userdefinelength]{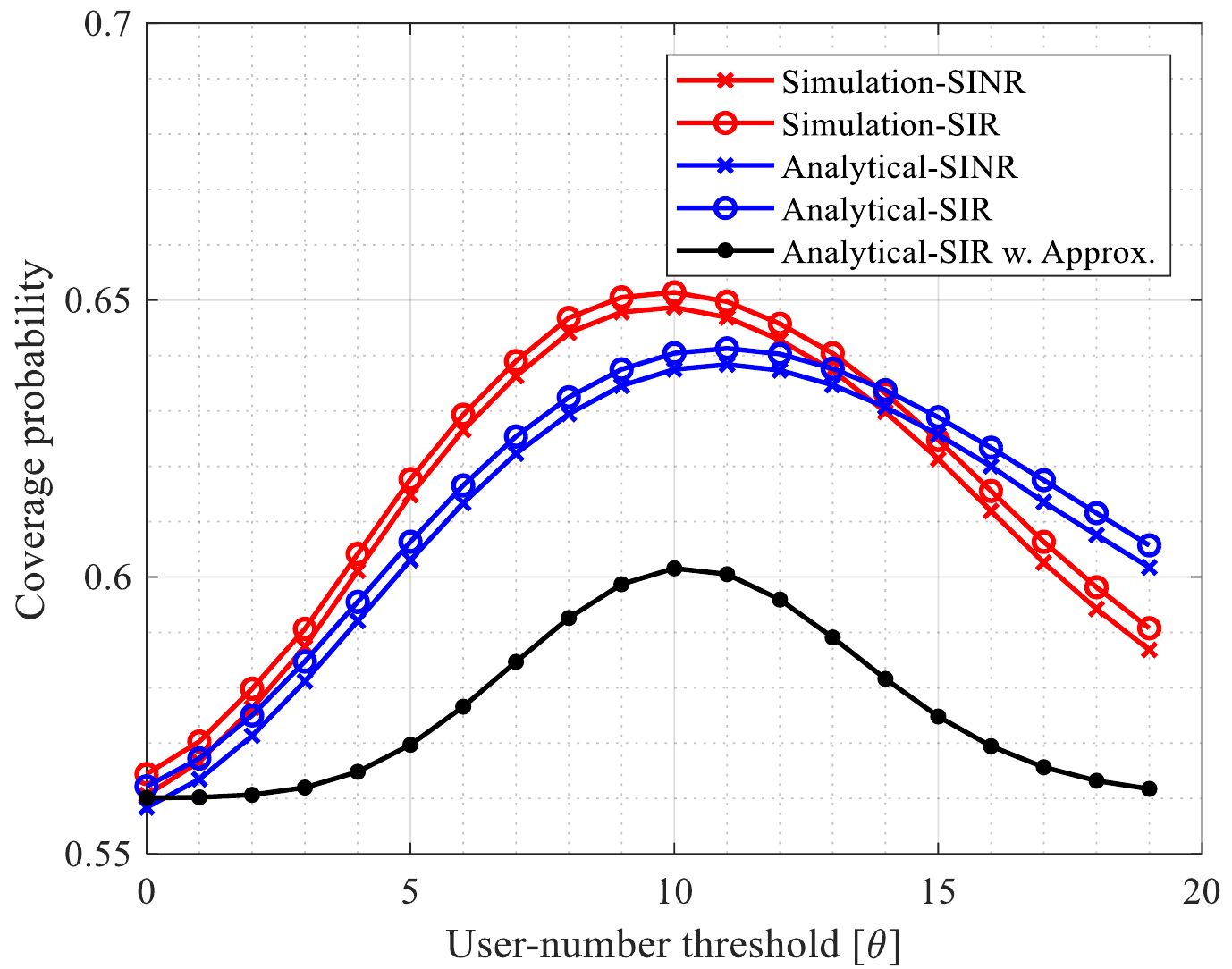}
  \captionof{figure}{Coverage probability for HomNet}
  \label{fig:cov_prob_homnet}
\end{figure}

Fig. \ref{fig:cov_prob_homnet} illustrates the coverage probability of the user-number threshold-based on/off control, comparing the analytical and simulated results. 
Equations \eqref{eq:Pc_alpha_is_4} and \eqref{eq:overall_Pc_inf_limited} are plotted as ``Analytical-SINR" and ``Analytical-SIR", respectively, in  Fig.~\ref{fig:cov_prob_homnet}. 
The same equation as \eqref{eq:overall_Pc_inf_limited}, with 
substituting $\tilde{p}_a(\theta,\gamma)$ and $\tilde{p}_1(\theta,\gamma)$ for $p_a(\theta,\gamma)$ and $p_1(\theta,\gamma)$  
respectively, is plotted as ``Analytical-SIR w. Approx.".

As shown in Fig. \ref{fig:cov_prob_homnet}, there is a slight difference between 
the analytical and simulated results. 
This is because \eqref{eq:overall_Pc_inf_limited} actually represents the approximated results, as discussed in Theorem~1. 
When we obtain the coverage probability conditioned on the event that the user is connected to the $n$-th closest BS, the condition that the BS should be active is mitigated to reduce complexity. 
Therefore, the simulated and the analytical results do not match exactly. 
The difference, however, is negligible, with a maximum of only 0.02.

Regarding the optimal $\theta$ for maximizing the coverage probability, 
the simulation result is in great agreement with the analytical result, that is, $\theta_{opt} \approx \lfloor \gamma + 0.5 \rfloor$. 
The maximum coverage probability is achieved with $\theta_{opt}=10$ for simulated results (both SIR and SINR) and ``Analytical-SIR w. Approx.," and it is achieved with $\theta_{opt}=11$ for ``Analytical-SIR." 
Consequently, despite the approximations to reduce complexity, the analysis results presented in this work are quite accurate.

\begin{figure}[!htb]
\centering
  \includegraphics[width=\userdefinelength]{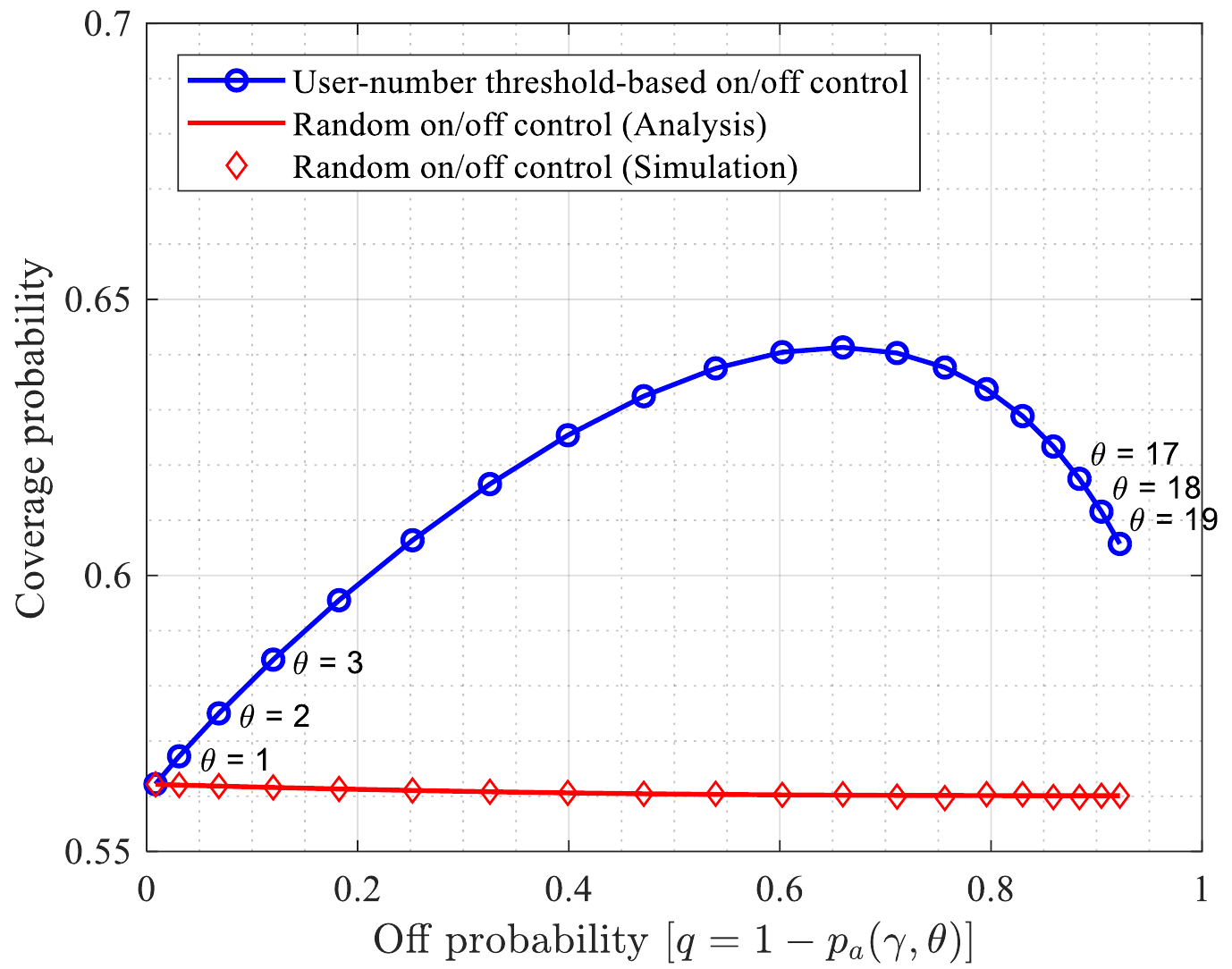}
  \caption{Coverage probability comparison between user-number threshold-based and random on/off control}
  \label{fig:cov_prob_comp}  
\end{figure}
\vspace{-9pt}

Fig. \ref{fig:cov_prob_comp} compares the random on/off control versus the 
user-number threshold-based on/off control with respect to the coverage probability. 
For a fair comparison, two schemes are compared with respect to the same off probability. 
The coverage probability of the user-number threshold-based on/off control is evaluated with $\theta$ increasing from 0 to 19. 
Then, the off probability of the user-number threshold-based on/off control is obtained as $1-p_a(\gamma, \theta)$. Next, the coverage probability of random control is evaluated according to $q = 1-p_a(\gamma, \theta)$. 
Note that for the random on/off control scheme, the simulation results are in great agreement with the analysis results in \eqref{eq:cov_prob_random}, and maintain an almost constant value. 

The result verified that the coverage probability could be enhanced by the user-number threshold-based on/off control compared to the baseline performance of random on/off control. 
Additional comparisons with other coordinated on/off controls are left for future work. This is because the main focus of this study is not to show the superior performance of the user-number threshold-based on/off control, but to provide essential  analysis tools.

\ifdefined\singlecolumn
\else
\begin{figure*}[!htb]
\centering
\begin{minipage}{.5\textwidth}
  \centering
  \includegraphics[width=\linewidth]{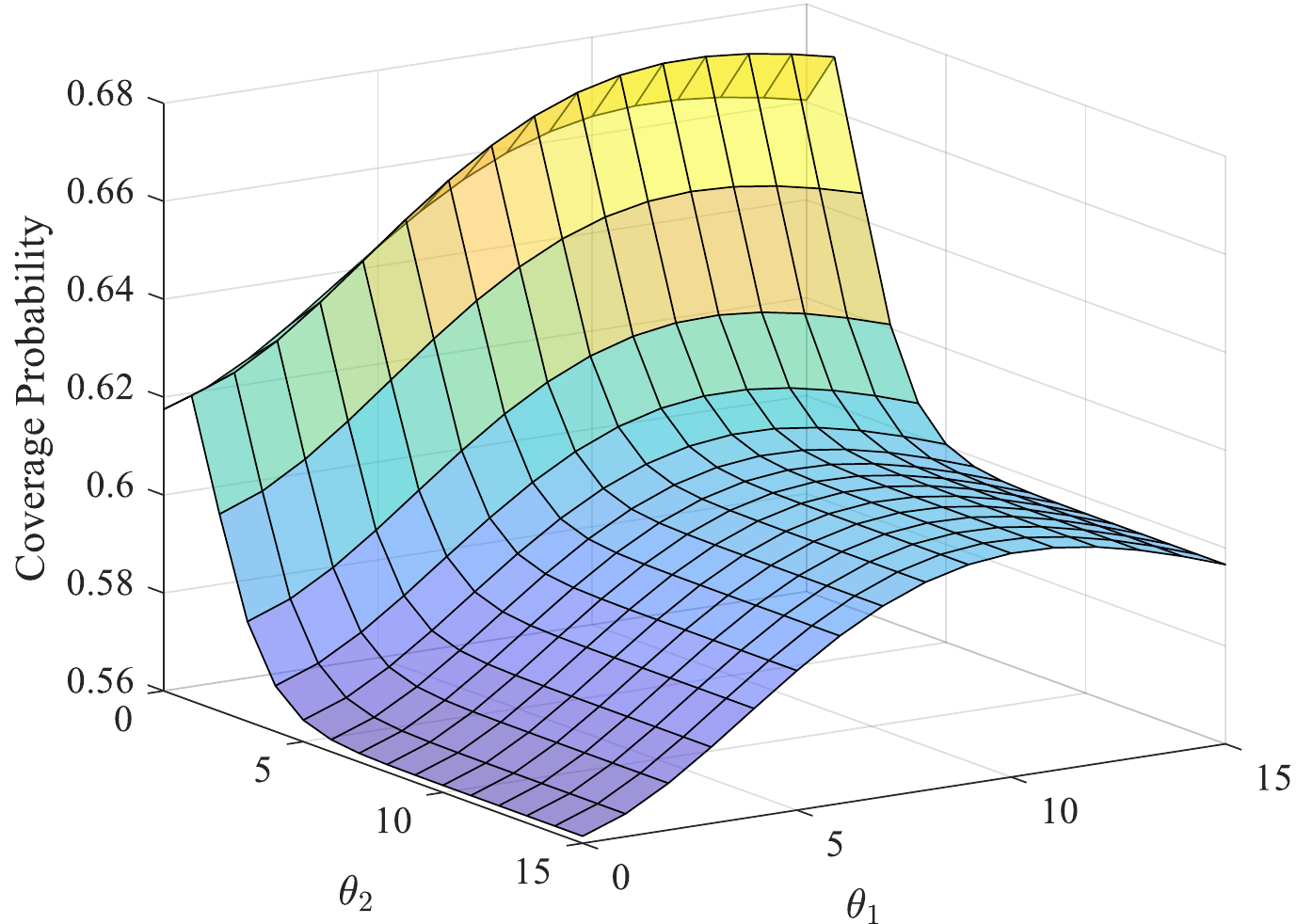}
  \captionof{figure}{Coverage probability for HetNet (Analysis)}
  \label{fig:HetNet_3D_Analysis}
\end{minipage}%
\begin{minipage}{.5\textwidth}
  \centering
  \includegraphics[width=\linewidth]{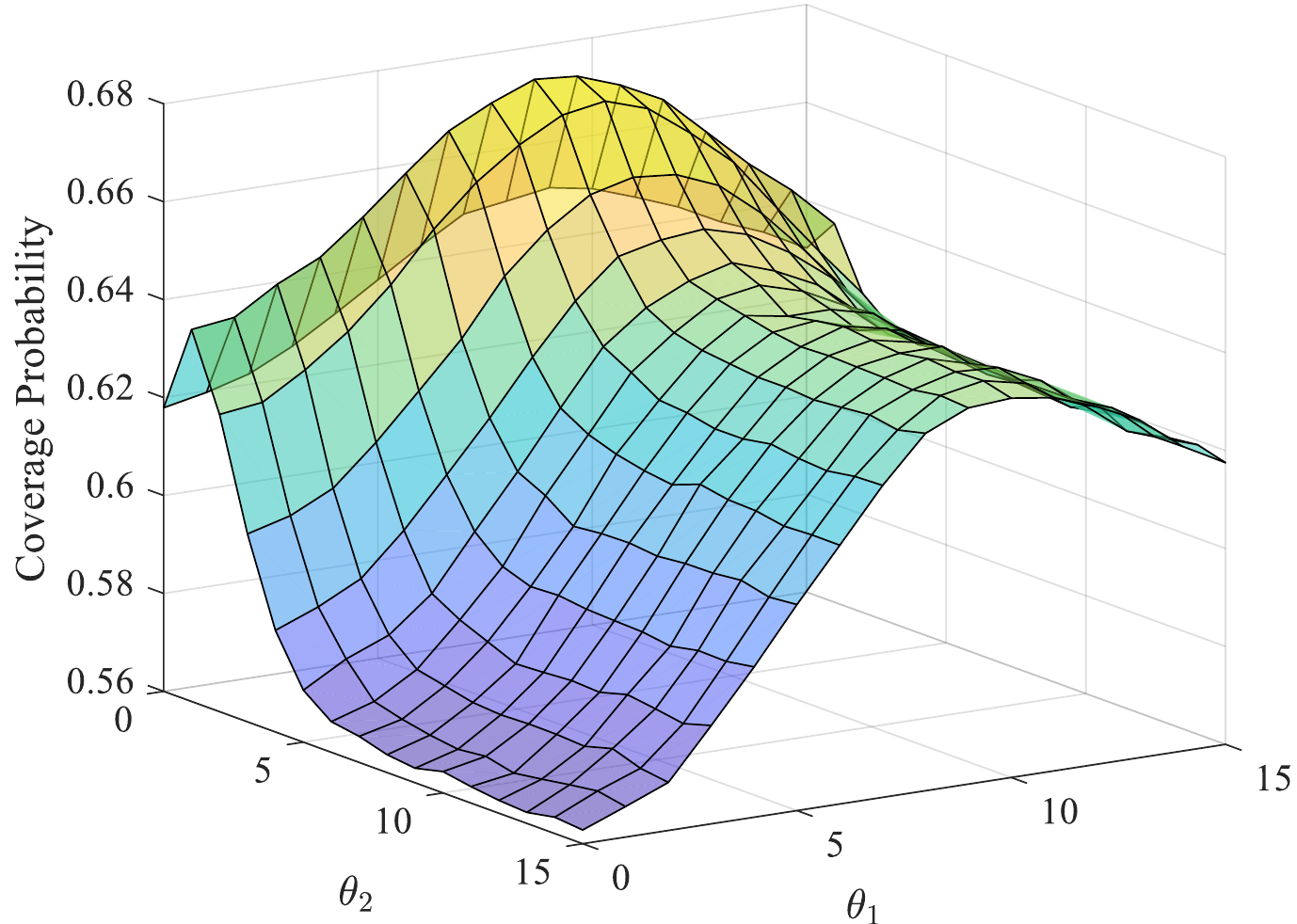}
  \captionof{figure}{Coverage probability for HetNet (Simulation)}
  \label{fig:HetNet_3D_Simulation}
\end{minipage}
\begin{minipage}{\textwidth}
\vspace{10pt}
\begin{minipage}{.5\textwidth}
  \centering
  \includegraphics[width=\linewidth]{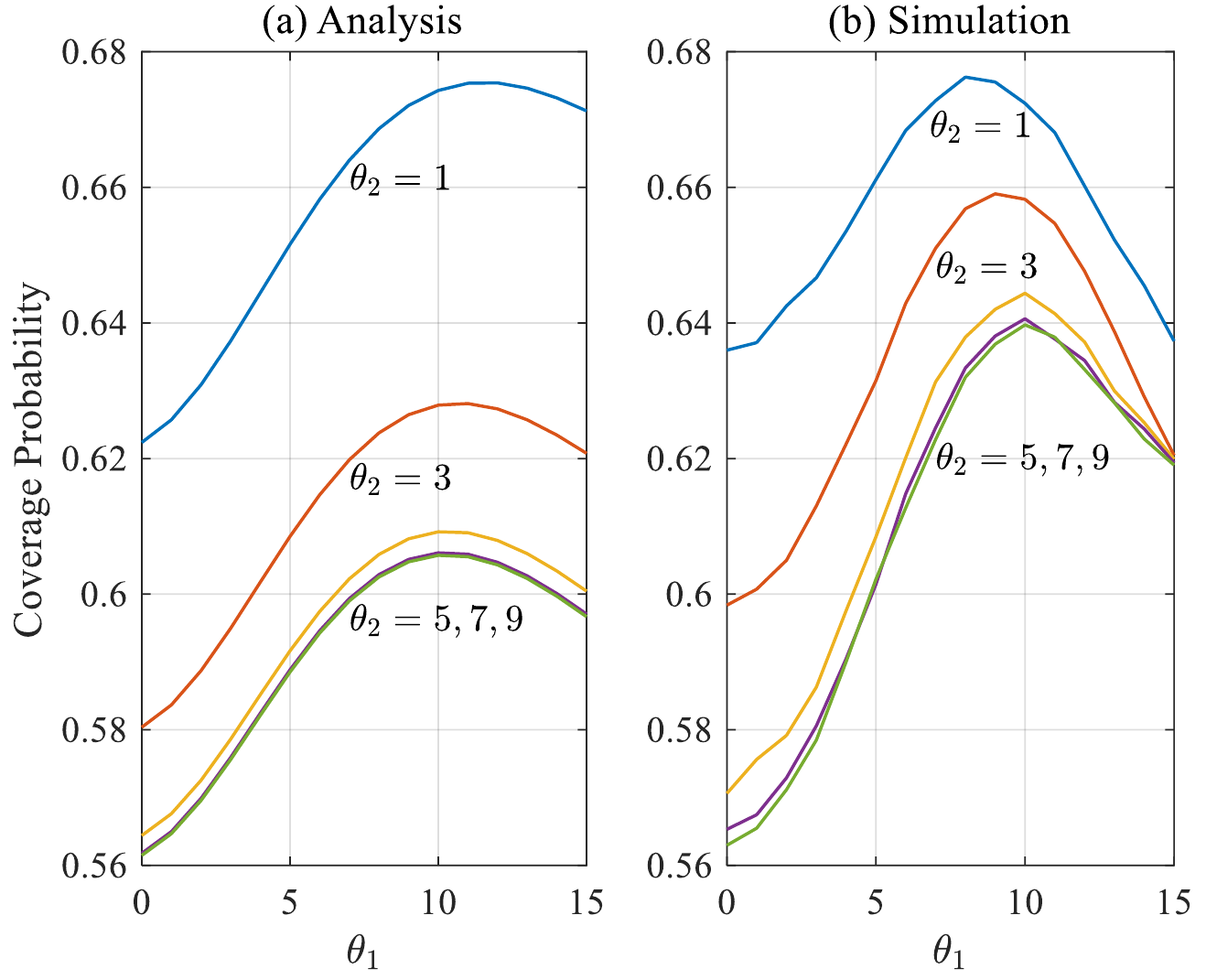}
  \captionof{figure}{Coverage probability for HetNet with respect to $\theta_1$}
  \label{fig:HetNet_Pc_wrt_theta1}
\end{minipage}%
\begin{minipage}{.5\textwidth}
  \centering
  \includegraphics[width=\linewidth]{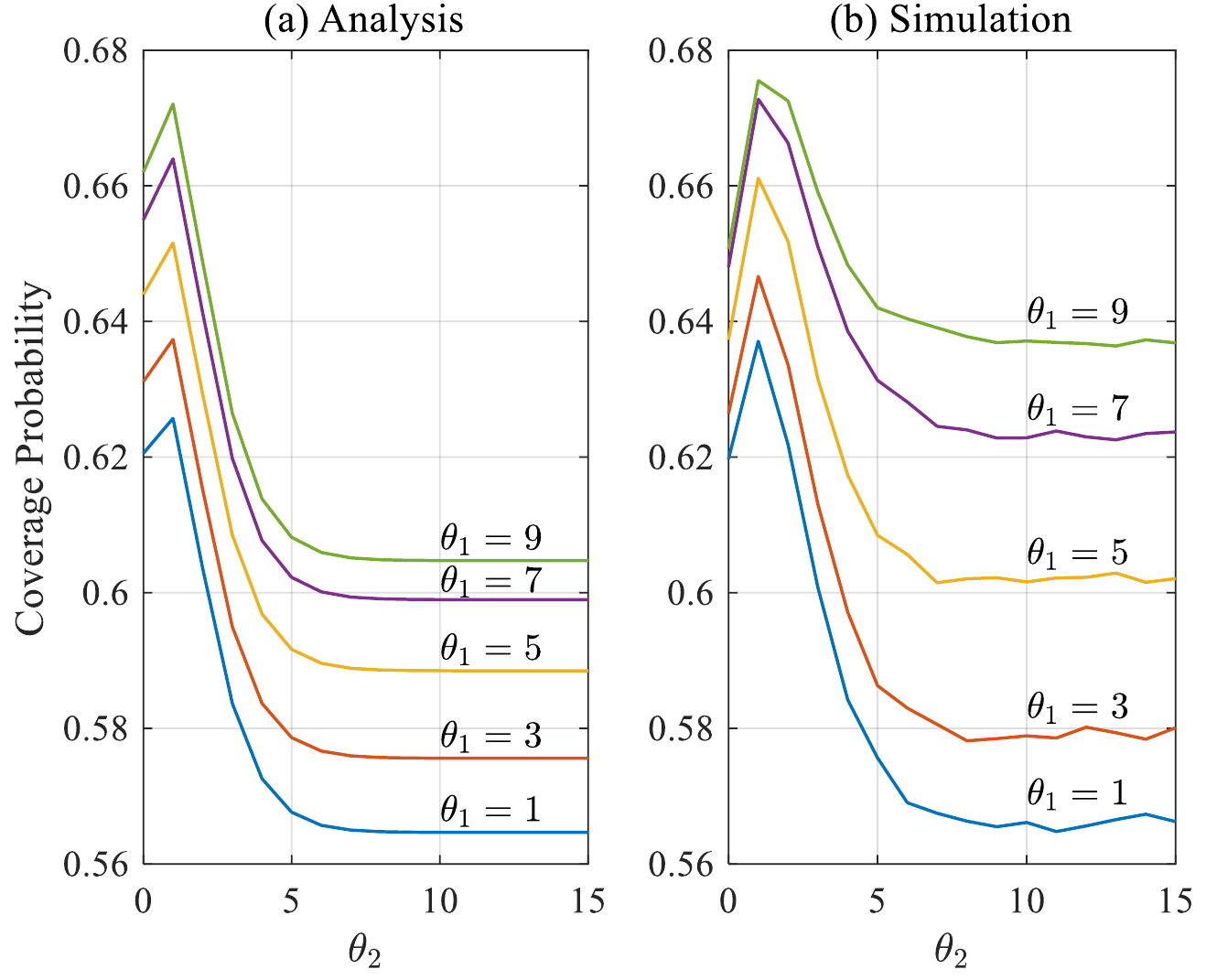}
  \captionof{figure}{Coverage probability for HetNet with respect to $\theta_2$}
  \label{fig:HetNet_Pc_wrt_theta2}
\end{minipage}
\end{minipage}
\end{figure*}
\fi

The HetNet simulations were conducted in a $2000~m \times 2000~m$ grid, with 500 trials. 
We assumed the HetNet had two tiers with different densities and powers. 
It was assumed that $\lambda_1 = \num{1e-4}$, $\lambda_2 = \num{1e-3}$, $\lambda_u = \num{20e-3}$, and $P_{t,1} = 10 P_{t,2}$, such that 
$\bar{\gamma}_1 = 10$ and $\bar{\gamma}_2 = 1$. 
As same for the HomNet, $T=0$~dB and $\alpha=4$. 
$\theta_1$ and $\theta_2$ vary from 0 to 14. 
Note that the scale of simulations, such as the size of the grid, the number of trials, and the range of variation of $\theta$ are reduced compared to the HomNet simulation, because of the limitation of computational time and resources. 

\ifdefined\singlecolumn
\begin{figure*}[!htbp]
\centering
\begin{minipage}{.5\textwidth}
  \centering
  \includegraphics[width=\linewidth]{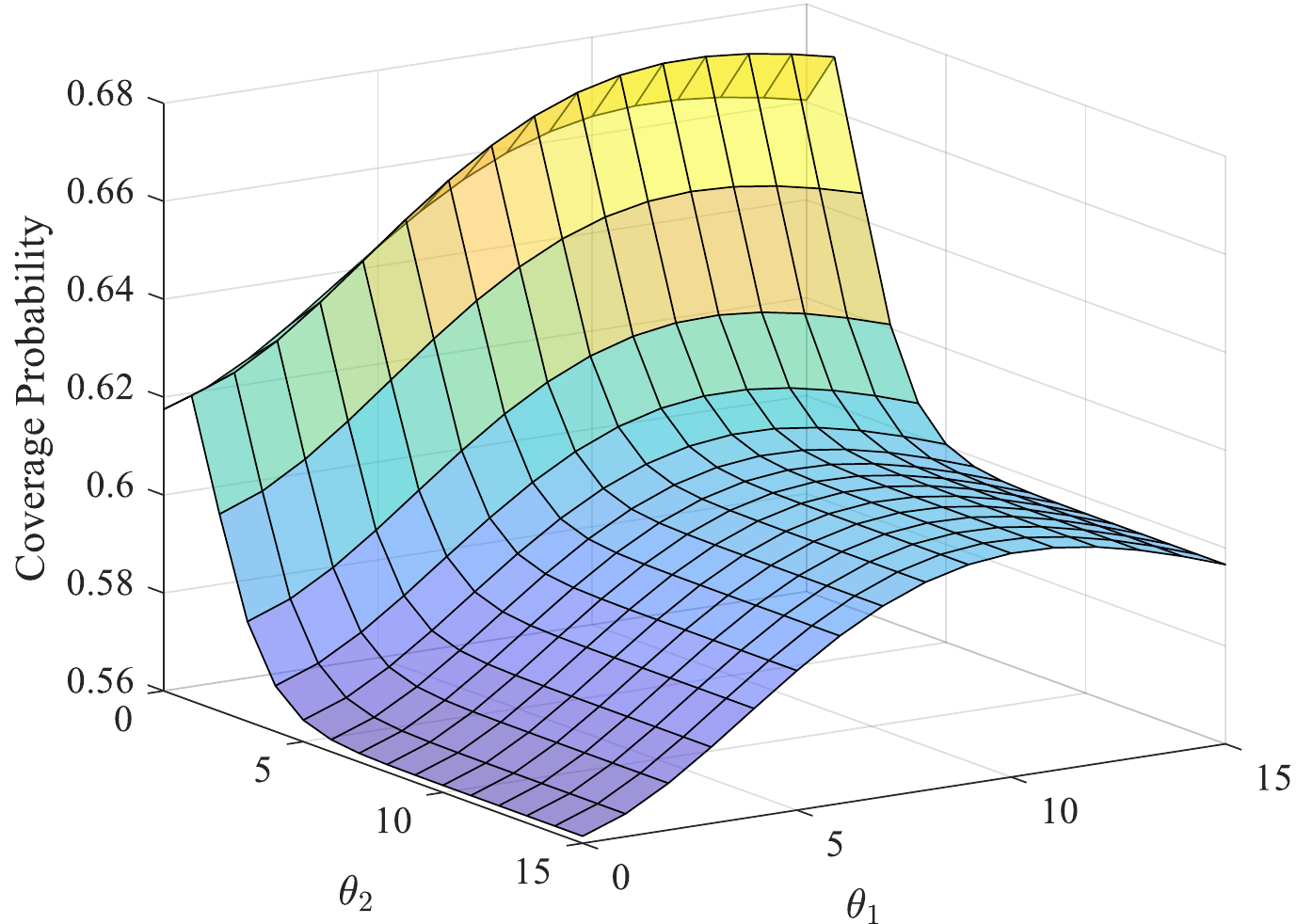}
  \captionof{figure}{Coverage probability for HetNet \\  (Analysis)}
  \label{fig:HetNet_3D_Analysis}
\end{minipage}%
\begin{minipage}{.5\textwidth}
  \centering
  \includegraphics[width=\linewidth]{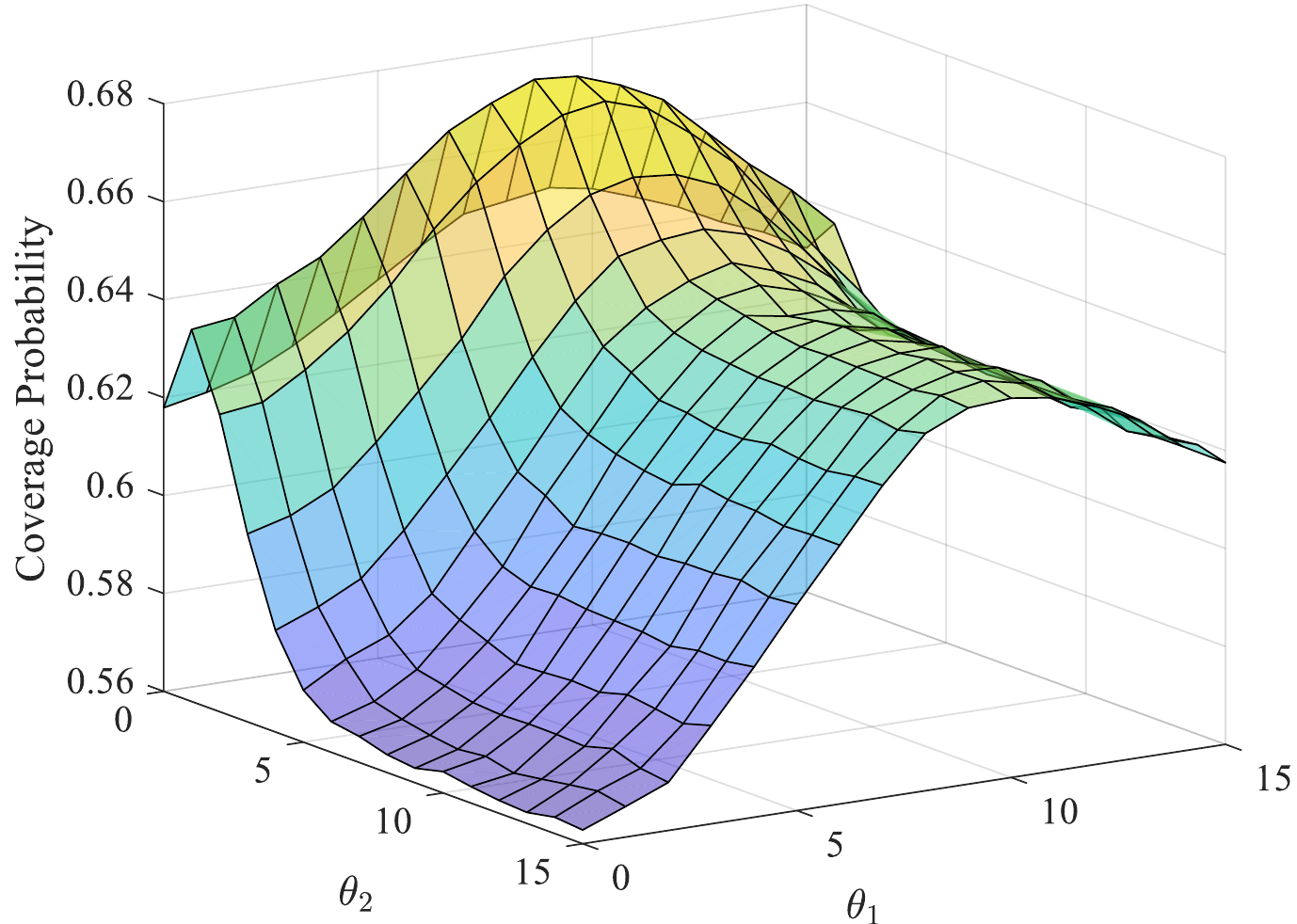}
  \captionof{figure}{Coverage probability for HetNet \\  (Simulation)}
  \label{fig:HetNet_3D_Simulation}
\end{minipage}
\begin{minipage}{\textwidth}
\vspace{10pt}
\begin{minipage}{.5\textwidth}
  \centering
  \includegraphics[width=\linewidth]{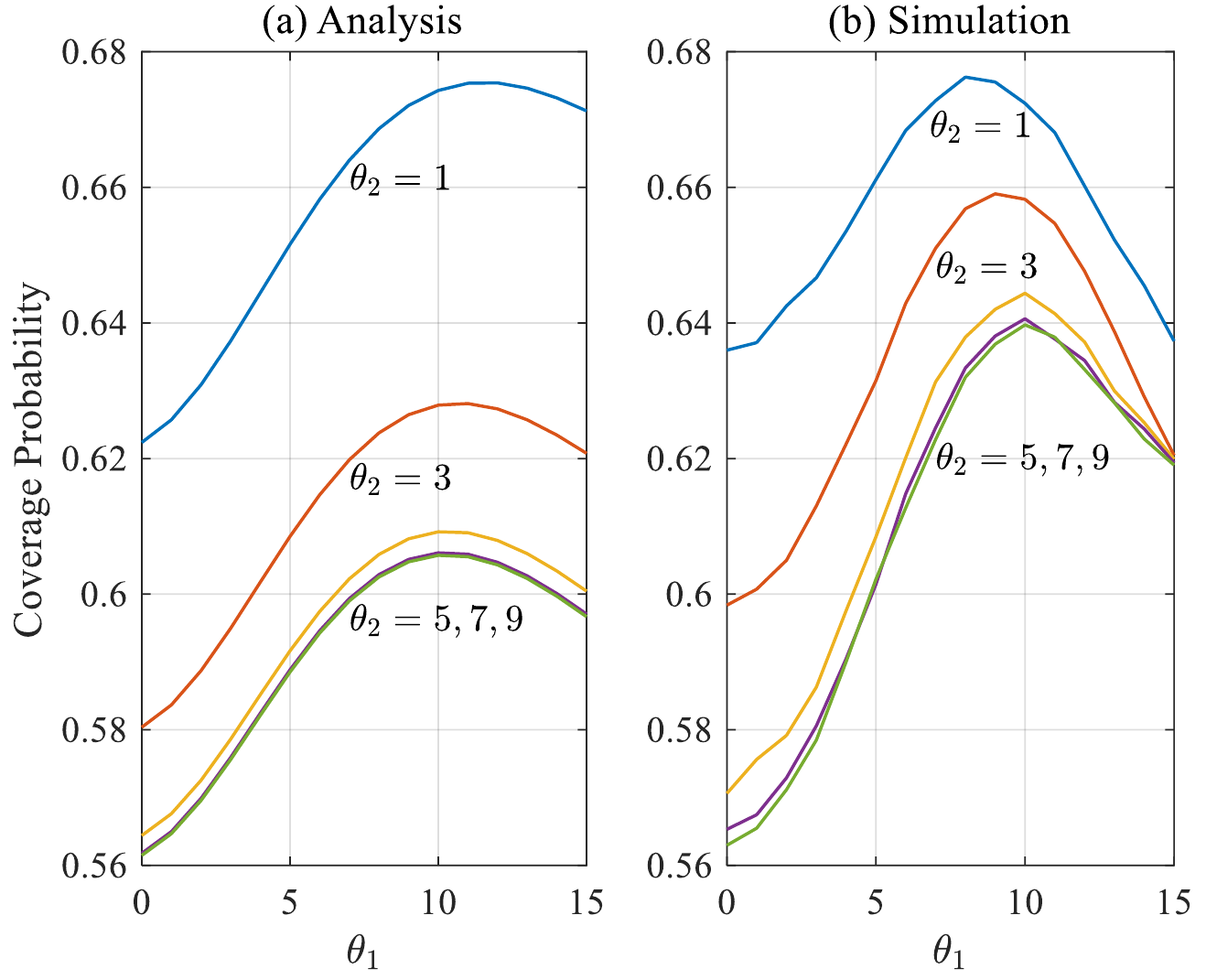}
  \captionof{figure}{Coverage probability for HetNet \\  with respect to $\theta_1$}
  \label{fig:HetNet_Pc_wrt_theta1}
\end{minipage}%
\begin{minipage}{.5\textwidth}
  \centering
  \includegraphics[width=\linewidth]{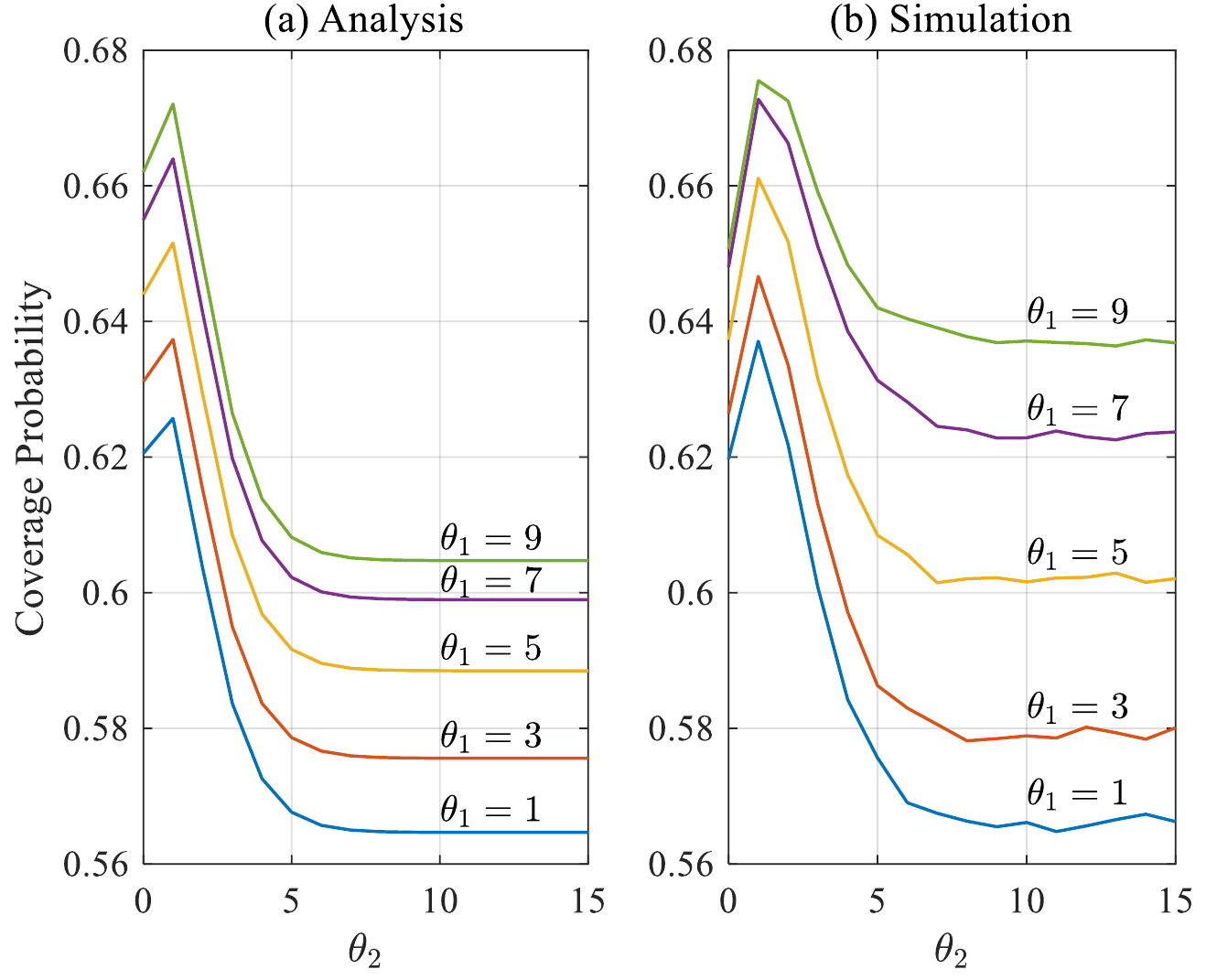}
  \captionof{figure}{Coverage probability for HetNet \\  with respect to $\theta_2$}
  \label{fig:HetNet_Pc_wrt_theta2}
\end{minipage}
\end{minipage}
\end{figure*}
\else
\fi

For the HetNet, the analytical result of \eqref{eq:pc_hetnet} is plotted in Fig. \ref{fig:HetNet_3D_Analysis}, and the simulation result is plotted in Fig. \ref{fig:HetNet_3D_Simulation}. 
Figs. \ref{fig:HetNet_Pc_wrt_theta1} and \ref{fig:HetNet_Pc_wrt_theta2} show two-dimensional plots on either side of $\theta_1$ and $\theta_2$ to clarify how the coverage probability for each threshold changes.
For the coverage probability, as with the HomNet, the simulation results of the HetNet are in great agreement with the analytical results. 
The difference between them is negligible, with a maximum of 0.04.

For the optimal user-number thresholds, we can approximate $\theta_{opt,1}$ and 
$\theta_{opt,2}$ as $\bar{\gamma}_1 = 10$ and $\bar{\gamma}_2 = 1$,  respectively. 
However, in Fig. 6, $\theta_{opt,1}$ shows a slight difference around 10 according to $\theta_2$.
This difference arises from the approximation (b) of (50), which is based on the assumption of a large $\bar{\gamma}_i~(\forall i)$. 
In the simulation setting, $\bar{\gamma}_1$ is large enough, at 10, but $\bar{\gamma}_2$ is small at 1. 
Although the result of the analysis does not exactly match the simulation result because of the approximation for simplicity, the difference is negligible. 


\section{Conclusion}
\label{sec:conclusion}
In this study, we analyzed the operation of a user-number threshold-based BS on/off control scheme, which turns off a BS when it has fewer users than a specific threshold called the user-number threshold. 
In particular, a spatial analysis of the BS on/off control system was conducted, to which a stochastic geometric approach was applied rather than the time-domain operation, which has been extensively studied in the literature. 
An approximated closed-form expression of coverage probability was derived when deploying the user-number threshold-based on/off control. 
In addition, the optimal user-number threshold for maximizing the coverage probability was derived analytically. 
Through the approximation of mitigating certain constraints, we obtained simplified and more tractable results with negligible errors. 
In addition to the HomNet, we derived the overall coverage probability and optimal user-number thresholds for the HetNet. 
These theoretical analyses demonstrate that the HetNet, for which analysis results are expected to be very complex, can be analyzed as a linear combination of HomNets with weighted densities. 
Furthermore, when the weighted density was large enough, it was shown that the optimal user-number threshold for each tier could be acquired independently of the other tiers in the network. 
Finally, we verified the theoretical contribution of this work by showing that the simulation results are highly consistent with the analytically derived results.

\ifdefined\singlecolumn
\else
\begin{figure*}[!hbt]
\normalsize
\setcounter{mytempeqncnt}{\value{equation}}
\setcounter{equation}{46}
\begin{align}
\label{eqn_dbl_x}
\mathbb{P}&(R_n > r, \mathbf{M} = \mathbf{m} , x_n \in \Phi_i ) =   \int_{r}^{\infty}\mbox{e}^{-\lambda_i \pi r_i^2}
\frac{2\left(\lambda_i \pi r_i^2 \right)^{m_i+1}}{r_i \Gamma(m_i+1)} 
\cdot \prod_{j=1/i}^{K} \left( \frac{1}{m_j!}\left( \pi \lambda_j \left(\frac{w_j}{w_i}r_i\right)^2  \right)^{m_j} \mbox{e}^{ - \pi \lambda_j \left(\frac{w_j}{w_i}r_i\right)^2 } \right) dr_i \nonumber \\
=& \frac{2 \left( \lambda_i \pi  \right)^{m_i+1}}{\Gamma(m_i+1)} \left( \frac{\pi}{w_j^2} \right)^{n-1-m_i} \prod_{j=1/i}^{K} \frac{\left( \lambda_j w_j^2 \right)^{m_j}}{m_j!}  
\int_{r}^{\infty} \exp \left(  -\pi \lambda_i \left( \frac{\sum_{j=1}^{K}\lambda_j w_j^2}{\lambda_i w_i^2} \right)  r_i^2 \right) r_i^{2n-1}  dr_i  \nonumber \\
\stackrel{(a)}{=}& \hspace{5pt}  \frac{q_i^n}{\Gamma(m_i+1) \left( \lambda_i w_i^2 \right)^{n-1-m_i}} \prod_{j=1/i}^{K} \frac{\left( \lambda_j w_j \right)^{m_j}}{m_j!} \int_{\frac{\pi \lambda_i r^2}{q_i}}^{\infty} e^{-t} t^{n-1} dt  
\nonumber \\
=& \hspace{5pt}  q_i \prod_{j=1}^{K} \frac{q_j^{m_j}}{m_j !} \int_{\frac{\pi \lambda_i r^2}{q_i}}^{\infty} e^{-t} t^{n-1} dt 
= q_i \prod_{j=1}^{K} \frac{q_j^{m_j}}{m_j !} \Gamma\left( n , \frac{\pi \lambda_i r^2}{q_i} \right)
= \hspace{5pt} \mathbb{P}\left(\mathbf{M}= \mathbf{m}  \right)  \sum_{k=0}^{n-1} e^{-\lambda_i \pi r^2 / q_i} \frac{(\lambda_i \pi r^2 / q_i)^k}{k!} q_i
\end{align}
\setcounter{equation}{\value{mytempeqncnt}}
\hrulefill
\vspace*{4pt}
\end{figure*}
\fi

\appendices
\section{Proof of Lemma~\ref{lem:pdf_of_dist_hetnet}}
\label{sec:lemma1}
Let $\mathbf{X}_{n-1}=\{x_1, \dots, x_{n-1}\}$ denote the set of BSs from the first to the $(n-1)$-th closest order, and $x_n$ denote the $n$-th closest BS. 
In the same process as in the case of HomNets, we obtain the probability that the distance of $x_n$ from the typical user located at the origin is greater than $r$. 

First, we obtain the CCDF of the distance from $x_n$ to the origin, assuming that each BS over $\mathbf{X}_{n-1}$ 
belongs to an arbitrary tier from one to $K$. 
Let the random variable of the number of BSs belonging to the $j$-th tier over $\mathbf{X}_{n-1}$ be $M_j$ ($\forall$~$j=1,\ldots,K$). 
Then, the set of random variables, that is,   $\mathbf{M}=\{M_1,\ldots,M_K\}$, follows a multinomial distribution with parameters $n-1$ and $\mathbf{q}$,
where $\mathbf{q}=\{q_1,\dots,q_K\}$.
The probability mass function of this multinomial distribution is given by
\ifdefined\singlecolumn
\be
\mathbb{P}\left(\mathbf{M} = \mathbf{m} \right)
= 
{ \displaystyle {n! \over m_1!\cdots m_K!}q_1^{m_1}\cdots q_K^{m_K}} 
=(n-1)! \displaystyle{ \prod_{i=1}^{K}  {\frac{q_i^{m_i}}{m_i!}} },
\ee
\else
\begin{align}
\mathbb{P}\left(\mathbf{M} = \mathbf{m} \right)
&= 
{ \displaystyle {n! \over m_1!\cdots m_K!}q_1^{m_1}\cdots q_K^{m_K}}  \nonumber \\
&=(n-1)! \displaystyle{ \prod_{i=1}^{K}  {\frac{q_i^{m_i}}{m_i!}} },
\end{align}
\fi
where $\mathbf{m}=\{m_1,\ldots,m_K\}$ and $\sum_{i=1}^k m_i=n-1 $. 

Now, we derive the joint probability that the distance from $x_n$ to the origin is greater than $r$, $\mathbf{M}$ follows arbitrary $\mathbf{m}$, and $x_n$ belongs to the $i$-th tier. 
It should be noted that there are $m_i$ BSs of the $i$-th tier over $\mathbf{X}_{n-1}$, and $x_n$ is the $(m_i+1)$-th closest BS because $x_n$ is assumed to belong to the $i$-th tier. 
Then, the joint probability is expressed as follows:
\ifdefined\singlecolumn
\be
\mathbb{P}(R_n > r , \mathbf{M} = \mathbf{m} , x_n \in \Phi_i ) = \int_{r}^{\infty} f_{R_{(m_{i}+1)}}  (r_i) \prod_{j=1/i}^{K} \mathbb{P}( N_j = m_j | r_i) dr_i,
\label{eq:hetNet_PDF_Rn}
\ee
\else
\begin{multline}
\mathbb{P}(R_n > r , \mathbf{M} = \mathbf{m} , x_n \in \Phi_i ) \\= \int_{r}^{\infty} f_{R_{(m_{i}+1)}}  (r_i) \prod_{j=1/i}^{K} \mathbb{P}( N_j = m_j | r_i) dr_i,
\label{eq:hetNet_PDF_Rn}
\end{multline}
\fi
where $f_{R_{(m_{i}+1)}}$ is the PDF of the distance to the $(m_i+1)$-th closest BS for the $i$-th tier in the same form of  \eqref{eq:pdf_of_dist_to_nth_BS}. 
$\mathbb{P}( N_j = m_j | r_i)$ is the probability that the number of BSs in the $j$-th tier over $\mathbf{X}_{n-1}$ is $m_j$. 
It is the probability that there are $m_j$ BSs in the circular area of radius $r_i w_j / w_i$. 
It is expressed as follows:
\be
\mathbb{P}( N_j = m_j | r_i) =  \frac{1}{m_j!}\left( \pi \lambda_i \left(\frac{w_j}{w_i}r_i\right)^2  \right)^{m_j} \mbox{e}^{ - \pi \lambda_j \left(\frac{w_j}{w_i}r_i\right)^2 }. 
\ee
Then, \eqref{eq:hetNet_PDF_Rn}
yields 
\ifdefined\singlecolumn
\begin{align}
\label{eqn_dbl_x}
\mathbb{P}&(R_n > r, \mathbf{M} = \mathbf{m} , x_n \in \Phi_i ) \nonumber \\ 
&= \int_{r}^{\infty}\mbox{e}^{-\lambda_i \pi r_i^2}
\frac{2\left(\lambda_i \pi r_i^2 \right)^{m_i+1}}{r_i \Gamma(m_i+1)} 
\cdot \prod_{j=1/i}^{K} \left( \frac{1}{m_j!}\left( \pi \lambda_j \left(\frac{w_j}{w_i}r_i\right)^2  \right)^{m_j} \mbox{e}^{ - \pi \lambda_j \left(\frac{w_j}{w_i}r_i\right)^2 } \right) dr_i \nonumber \\
&= \frac{2 \left( \lambda_i \pi  \right)^{m_i+1}}{\Gamma(m_i+1)} \left( \frac{\pi}{w_j^2} \right)^{n-1-m_i} \prod_{j=1/i}^{K} \frac{\left( \lambda_j w_j^2 \right)^{m_j}}{m_j!}  
\int_{r}^{\infty} \exp \left(  -\pi \lambda_i \left( \frac{\sum_{j=1}^{K}\lambda_j w_j^2}{\lambda_i w_i^2} \right)  r_i^2 \right) r_i^{2n-1}  dr_i  \nonumber \\
&\stackrel{(a)}{=} \frac{q_i^n}{\Gamma(m_i+1) \left( \lambda_i w_i^2 \right)^{n-1-m_i}} \prod_{j=1/i}^{K} \frac{\left( \lambda_j w_j \right)^{m_j}}{m_j!} \int_{\frac{\pi \lambda_i r^2}{q_i}}^{\infty} e^{-t} t^{n-1} dt  
 \\
&= q_i \prod_{j=1}^{K} \frac{q_j^{m_j}}{m_j !} \int_{\frac{\pi \lambda_i r^2}{q_i}}^{\infty} e^{-t} t^{n-1} dt 
= q_i \prod_{j=1}^{K} \frac{q_j^{m_j}}{m_j !} \Gamma\left( n , \frac{\pi \lambda_i r^2}{q_i} \right) \nonumber \\
&=  \mathbb{P}\left(\mathbf{M}= \mathbf{m}  \right)  \sum_{k=0}^{n-1} e^{-\lambda_i \pi r^2 / q_i} \frac{(\lambda_i \pi r^2 / q_i)^k}{k!} q_i, \nonumber
\end{align}
where (a) follows from the definition of $q_i$ in \eqref{eq:prob_of_tier} with substituting $\pi \lambda_i / q_i r_i^2$ with $t$.
\else
(47), where (a) follows from the definition of $q_i$ in \eqref{eq:prob_of_tier} with substituting $\pi \lambda_i / q_i r_i^2$ with $t$.
\setcounter{equation}{47}
\fi

Consequently, 
\ifdefined\singlecolumn
\begin{align}
\mathbb{P}(R_n > r  &| x_n \in \Phi_i ) 
= \sum_{\mathbf{m} \in \mbox{All}} \mathbb{P}(R_n > r , \mathbf{M} = \mathbf{m} | x_n \in \Phi_i ) \nonumber \\
&= \sum_{\mathbf{m} \in \mbox{All}} \frac{\mathbb{P}(R_n > r , \mathbf{M} = \mathbf{m} , x_n \in \Phi_i )}{q_i} 
= \sum_{k=0}^{n-1} e^{-\lambda_i \pi r^2 / q_i} \frac{(\lambda_i \pi r^2 / q_i)^k}{k!},
\end{align}
\else
\begin{align}
\mathbb{P}(R_n > r  &| x_n \in \Phi_i ) \nonumber \\
&= \sum_{\mathbf{m} \in \mbox{All}} \mathbb{P}(R_n > r , \mathbf{M} = \mathbf{m} | x_n \in \Phi_i ) \nonumber \\
&= \sum_{\mathbf{m} \in \mbox{All}} \frac{\mathbb{P}(R_n > r , \mathbf{M} = \mathbf{m} , x_n \in \Phi_i )}{q_i} \nonumber 
\\
&= \sum_{k=0}^{n-1} e^{-\lambda_i \pi r^2 / q_i} \frac{(\lambda_i \pi r^2 / q_i)^k}{k!},
\end{align}
\fi
where $\mathbb{P}( x_n \in \Phi_i ) = q_i$, and $\sum_{\mathbf{m} \in \mbox{All}} \mathbb{P} (\mathbf{M}=\mathbf{m}) =  1$.
Then, the CDF is $F_{R_{n|i}} (r) =  1 - \mathbb{P}(R_n \geq r  | x_n \in \Phi_i ) $, and its PDF
is given by \eqref{eq:pdf_of_dist_hetnet}.

\section{Proof of Lemma~\ref{lem:laplace_tf_jth_intf}}
\label{sec:lemma2}

The total interference is the sum of the interferences at all tiers from one to $K$. Thus, the Laplace transform of the total interference is the product of the Laplace transforms of interferences at all tiers, expressed as follows: 
\begin{align}
\mathbb{E}_{I_r} \left[  \exp \left( - \frac{Tr^\alpha}{P_{t,i}} I_r  \right) \right] =
\prod_{j=1}^{K} \mathbb{E}_{I_j} \left[  \exp \left(  - s I_j \right) \right],
\label{eq:def_laplace_tf_jth_intf}
\end{align}
where $I_j$ is the interference power from the $j$-th tier.

From the known results regarding 
the probability generating functional (PGFL) of PPPs in \cite{Jeffrey:2011},
the Laplace transform of the interference power from the $j$-th tier is expressed as
\ifdefined\singlecolumn
\begin{align}
\mathbb{E}_{I_{j}} &\left[ \exp\left( -\frac{T r^\alpha}{P_{t,i}} I_j \right)  \right] \nonumber \\
&= \exp \Biggl( -2 \pi p_{a,j}(\theta_j, \bar{\gamma}_j ) \lambda_j \left. \cdot {\displaystyle \int_{r_j=r \left(  \frac{w_i}{w_j} \right)}^{\infty} \left( 1- \frac{1}{1+T\frac{P_{t,j}}{P_{t,i}}r^{\alpha}v^{-\alpha}} v dv  \right) }\right) \nonumber \\
&= \exp \left( -\pi r^2 p_{a,j}(\theta_j, \bar{\gamma}_j ) \frac{P_{t,j}^{1/\alpha}}{P_{t,i}^{1/\alpha}}
 T^{2/\alpha} { \displaystyle \int_{T^{-2/\alpha}}^{\infty}  \frac{1}{1+u^{\alpha/2}} du }
\right) \nonumber \\
&= \exp \left( -\pi r^2 p_{a,j}(\theta_j, \bar{\gamma}_j ) \lambda_j \frac{w_j^2}{w_i^2} \rho (T,\alpha)   \right). 
\end{align}
\else
\begin{align}
&\quad \mathbb{E}_{I_{j}} \left[ \exp\left( -\frac{T r^\alpha}{P_{t,i}} I_j \right)  \right] = \exp \Biggl( -2 \pi p_{a,j}(\theta_j, \bar{\gamma}_j ) \lambda_j \nonumber  \\
&\hspace{50pt}\left. \cdot {\displaystyle \int_{r_j=r \left(  \frac{w_i}{w_j} \right)}^{\infty} \left( 1- \frac{1}{1+T\frac{P_{t,j}}{P_{t,i}}r^{\alpha}v^{-\alpha}} v dv  \right) }\right) \nonumber \\
&= \exp \left( -\pi r^2 p_{a,j}(\theta_j, \bar{\gamma}_j ) \frac{P_{t,j}^{1/\alpha}}{P_{t,i}^{1/\alpha}}
 T^{2/\alpha} { \displaystyle \int_{T^{-2/\alpha}}^{\infty}  \frac{1}{1+u^{\alpha/2}} du }
\right) \nonumber \\
&= \exp \left( -\pi r^2 p_{a,j}(\theta_j, \bar{\gamma}_j ) \lambda_j \frac{w_j^2}{w_i^2} \rho (T,\alpha)   \right). 
\end{align}
\fi

Then, \eqref{eq:def_laplace_tf_jth_intf} is further expressed as
\ifdefined\singlecolumn
\begin{align}
\mathbb{E}_{I_r} &\left[  \exp \left( - \frac{Tr^\alpha}{P_{t,i}} I_r  \right) \right] \nonumber \\
&= \exp{ \left( - \pi r^2 \frac{\sum_{j=1}^{K} p_{a,j}(\theta_j, \bar{\gamma}_j ) \lambda_j w_j^2}{w_i^2} \rho(T,\alpha) \right) } \nonumber \\
&= \exp \left( - \pi r^2 \frac{\lambda_i}{q_i} \frac{\sum_{j=1}^{K} p_{a,j}(\theta_j, \bar{\gamma}_j ) \lambda_j w_j^2}{\sum_{j=1}^{K} \lambda_j w_j^2} \rho(T,\alpha)  \right) \nonumber \\
&= \exp \left( - \pi r^2 \bar{\lambda}_i \sum_{j=1}^{K} p_{a,j}(\theta_j, \bar{\gamma}_j )  q_j \rho(T,\alpha)  \right) 
\nonumber \\
&= \exp \biggl( - \pi r^2 \bar{\lambda}_i  p^{avr}_a(\{\theta\},\{\bar{\gamma} \}) \rho(T,\alpha) \biggr).
\end{align}
\else
\begin{align}
&\mathbb{E}_{I_r} \left[  \exp \left( - \frac{Tr^\alpha}{P_{t,i}} I_r  \right) \right] \nonumber \\
&= \exp{ \left( - \pi r^2 \frac{\sum_{j=1}^{K} p_{a,j}(\theta_j, \bar{\gamma}_j ) \lambda_j w_j^2}{w_i^2} \rho(T,\alpha) \right) } \nonumber \\
&= \exp \left( - \pi r^2 \frac{\lambda_i}{q_i} \frac{\sum_{j=1}^{K} p_{a,j}(\theta_j, \bar{\gamma}_j ) \lambda_j w_j^2}{\sum_{j=1}^{K} \lambda_j w_j^2} \rho(T,\alpha)  \right) \nonumber \\
&= \exp \left( - \pi r^2 \bar{\lambda}_i \sum_{j=1}^{K} p_{a,j}(\theta_j, \bar{\gamma}_j )  q_j \rho(T,\alpha)  \right) \nonumber \\
&= \exp \biggl( - \pi r^2 \bar{\lambda}_i  p^{avr}_a(\{\theta\},\{\bar{\gamma} \}) \rho(T,\alpha) \biggr).
\end{align}
\fi

\bibliographystyle{IEEEtran}

\end{document}